\documentclass[sigconf]{acmart}

\usepackage{diagbox}
\usepackage{tabularx}
\usepackage{amsmath}
\usepackage{algorithm}
\usepackage{algpseudocode}
\usepackage{mathrsfs}
\usepackage{amsfonts}
\usepackage{dutchcal}
\usepackage{multirow}
\usepackage{subfigure}
\usepackage{algorithm}
\usepackage{algorithmicx}
\usepackage{amsmath,amsfonts}
\usepackage{array}
\usepackage{balance}
 \usepackage{booktabs}
\usepackage[skip=1pt,labelfont=bf]{caption}
 \usepackage{calligra}
\usepackage{color}
\usepackage{colortbl}
\usepackage{courier}
\usepackage{csvsimple}
\usepackage{enumitem}
\usepackage{fancybox}
\usepackage{fontenc}
\usepackage{graphicx}
\usepackage{listings}
\usepackage{longtable}
\usepackage{lscape}
\usepackage{makecell}
\usepackage{marvosym}
\usepackage{moreverb}
\usepackage{multicol}
\usepackage{multirow}
\usepackage{makecell}
\usepackage{pifont}
\usepackage{rotating}
\usepackage{setspace}
\usepackage[most]{tcolorbox}
\usepackage{threeparttable}
\usepackage{tikz}
\usepackage{soul}
\usepackage[normalem]{ulem}
\usepackage{url}
\usepackage{soul}
\usepackage{wasysym}
\usepackage{xspace}
\usepackage{amsthm}

\algnewcommand\algorithmicforeach{\textbf{for each}}
\algdef{S}[FOR]{ForEach}[1]{\algorithmicforeach\ #1\ \algorithmicdo}

\newcolumntype{L}[1]{>{\raggedright\let\newline\\\arraybackslash\hspace{0pt}}m{#1}}
\newcolumntype{C}[1]{>{\centering\let\newline\\\arraybackslash\hspace{0pt}}m{#1}}
\newcolumntype{R}[1]{>{\raggedleft\let\newline\\\arraybackslash\hspace{0pt}}m{#1}}

\definecolor{codegreen}{rgb}{0,0.6,0}
\definecolor{codered}{rgb}{1,0,0}
\definecolor{codegray}{rgb}{0.9,0.9,0.9}
\definecolor{codepurple}{rgb}{0.58,0,0.82}
\definecolor{backcolour}{rgb}{0.95,0.95,0.92}
\definecolor{lightgray}{gray}{0.9}

\lstdefinestyle{mystyle}{
    commentstyle=\color{codegreen},
    keywordstyle=\color{magenta},
    numberstyle=\small\color{black},
    stringstyle=\color{codepurple},
    basicstyle=\scriptsize\ttfamily,
    breakatwhitespace=false,
    breaklines=true,
    captionpos=b,
    keepspaces=true,
    showspaces=false,
    showstringspaces=false,
    showtabs=false,
    tabsize=2
}

\lstdefinelanguage{diff}{
  morecomment=[f][\color{blue}]{@@},     
  morecomment=[f][\color{red}]-,         
  morecomment=[f][\color{codegreen}]+,       
  morecomment=[f][\color{red}]{---}, 
  morecomment=[f][\color{codegreen}]{+++},
}

\setlist{noitemsep} 

\definecolor{darkpastelred}{rgb}{0.76, 0.23, 0.13}
\definecolor{ao(english)}{rgb}{0.0, 0.5, 0.0}

\definecolor{darkpastelred}{rgb}{0.76, 0.23, 0.13}
\definecolor{ao(english)}{rgb}{0.0, 0.5, 0.0}

\definecolor{yellow}{RGB}{255,255,153}
\definecolor{grey}{RGB}{224,224,224}

\newboolean{showcomments}
\setboolean{showcomments}{true}
\ifthenelse{\boolean{showcomments}}
 { \newcommand{\mynote}[2]{
      \fbox{\bfseries\sffamily\scriptsize#1}
        {\small$\blacktriangleright$\textsf{\emph{#2}}$\blacktriangleleft$}}}
        { \newcommand{\mynote}[2]{}}

\definecolor{DarkOrange}{rgb}{0.8,0.3,0.0}
\definecolor{DarkCyan}{rgb}{0.0, 0.55, 0.55}
\definecolor{DarkCyel}{rgb}{1.0, 0.49, 0.0}
\definecolor{yellow-green}{rgb}{0.6, 0.8, 0.2}

\newcolumntype{?}{!{\vrule width 1pt}}

\newcommand{\etal}{\emph{et~al.}\xspace}
\newcommand*{\ie}{i.e., }
\newcommand*{\eg}{e.g., }

\newcommand{\tool}{Patcherizer\xspace}

\newcommand{\find}[1]{
\begin{tcolorbox}[leftrule=0.2mm,toprule=0mm,bottomrule=0mm,left=0.0pt,right=0pt,top=0pt,bottom=0pt]
\em #1
\end{tcolorbox}
}

\AtBeginDocument{%
  \providecommand\BibTeX{{%
    \normalfont B\kern-0.5em{\scshape i\kern-0.25em b}\kern-0.8em\TeX}}}

\begin{document}
\title{Learning to Represent Patches}

\author{Xunzhu Tang}
\email{xunzhu.tang@uni.lu}
\affiliation{%
  \institution{University of Luxembourg}
 	\country{Luxembourg}
}

\author{Haoye Tian}\authornote{Corresponding author.}
\email{haoye.tian@uni.lu}
\affiliation{%
  \institution{University of Luxembourg}
 	\country{Luxembourg}
}

\author{Zhenghan Chen}
\email{1979282882@pku.edu.cn}
\affiliation{%
  \institution{Peking University}
  \country{China}
}
\author{Weiguo Pian}
\email{weiguo.pian@uni.lu}
\affiliation{%
  \institution{University of Luxembourg}
 	\country{Luxembourg}
}
\author{Saad Ezzini}
\email{s.ezzini@lancaster.ac.uk}
\affiliation{%
  \institution{Lancaster University}
 	\country{United Kingdom}
}

\author{Abdoul Kader Kaboré}
\email{abdoulkader.kabore@uni.lu}
\affiliation{%
  \institution{University of Luxembourg}
  \country{Luxembourg}
}

\author{Andrew Habib}
\email{andrew.a.habib@gmail.com}
\affiliation{%
  \institution{University of Luxembourg}
 	\country{Luxembourg}
}

\author{Jacques Klein}
\email{jacques.klein@uni.lu}
\affiliation{%
  \institution{University of Luxembourg}
 	\country{Luxembourg}
}

\author{Tegawendé F. Bissyandé}
\email{tegawende.bissyande@uni.lu}
\affiliation{%
  \institution{University of Luxembourg}
 	\country{Luxembourg}
}


\begin{abstract}
Patch representation plays a pivotal role in automating numerous software engineering tasks, such as classifying patch correctness or generating natural language summaries of code changes. Recent studies have leveraged deep learning to derive effective patch representation, primarily focusing on capturing changes in token sequences or Abstract Syntax Trees (ASTs). However, these current state-of-the-art representations do not explicitly calculate the intention semantic induced by the change on the AST, nor do they optimally explore the surrounding contextual information of the modified lines.
To address this, we propose a new patch representation methodology, \tool, which we refer to as our tool. This innovative approach explores the intention features of the context and structure, combining the context around the code change along with two novel representations. These new representations capture the sequence intention inside the code changes in the patch and the graph intention inside the structural changes of AST graphs before and after the patch. This comprehensive representation allows us to better capture the intentions underlying a patch.
\tool builds on graph convolutional neural networks for the structural input representation of the intention graph and on transformers for the intention sequence representation.  
We assess the generalizability of \tool's learned embeddings on three tasks: (1) Generating patch description in NL, (2) Predicting patch correctness in program repair, 
and (3) Patch intention detection.
Experimental results show that the learned patch representation is effective for all three tasks and achieves superior performance to SOTA approaches. 
For instance, on the popular task of patch description generation (a.k.a. commit message generation), \tool achieves an average improvement of 19.39\%, 8.71\%, and 34.03\% in terms of BLEU,
ROUGE-L, and METEOR metrics, respectively. 
\end{abstract}

\maketitle

\section{Introduction}\label{sec:intro}

A software patch represents the source code differences between two software versions. It has a dual role: on the one hand, it serves as a formal summary of the code changes that a developer intends to make on a code base; on the other hand, it is used as the main input specification for automating software evolution. Patches are thus a key artifact that is pervasive across the software development life cycle. In recent years, building on empirical findings on the repetitiveness of code changes~\cite{barr2014plastic},  several approaches have built machine learning models based on patch datasets to automate various software engineering tasks such as patch description generation~\cite{linares2015changescribe,buse2010automatically,cortes2014automatically,jiang2017automatically,xu2019commit,liu2019generating,liu2018neural,liu2020atom}, code completion~\cite{svyatkovskiy2019pythia,liu2020multi,liu2020self,ciniselli2021empirical,pian2022metatptrans}, patch correctness assessment \cite{tian2022change}, and just-in-time defect prediction \cite{hoang2019deepjit,kamei2016studying}.

Early work manually crafted a set of hand-picked features to represent a patch~\cite{kamei2016studying,kamei2012large}. 
Recently, the successful application of deep learning to learn powerful representations of text, signal, and images~\cite{devlin2018bert,niemeyer2021giraffe,qin2021co,li2021attention,tang2021moto,pian2023metatptrans,wang2022hienet} led to adopting similar techniques in software engineering where researchers develop deep ML models to represent code and code changes~\cite{yin2018learning,hoang2020cc2vec,feng2020codebert,nie2021coregen,jiang2021cure,tian2022change,pian2022metatptrans,liu2023ccrep}. 
Initially, these approaches treated code~\cite{feng2020codebert,elnaggar2021codetrans,DBLP:conf/emnlp/0034WJH21,DBLP:conf/iclr/GuoRLFT0ZDSFTDC21,DBLP:conf/iclr/GuoRLFT0ZDSFTDC21} and other code-like artifacts, such as patches~\cite{xu2019commit,nie2021coregen,dong2022fira,liu2020atom}, as a sequence of tokens and thus employ natural language processing methods to extract code in text format. 
But because source code is not just text, researchers noticed the importance of code structure and began to use the code structure -- e.g. through Abstract Syntax Trees (ASTs) -- to capture the underlying structural information in source code~\cite{zhang2019novel,Alon2019,alonanycodegen,DBLP:conf/iclr/GuoRLFT0ZDSFTDC21}.
However, the differences in code token sequences (usually represented by {\tt +} and {\tt -} lines in the textual diff format) are not be sufficient to represent the full semantics of the code change as the {\tt +} and {\tt -} symbols do not have an inherent meaning that a DL model can learn.
Therefore, recent work such as commit2vec~\cite{cabrera2021commit2vec}, C\textsuperscript{3}~\cite{brody2020structural}, and CC2Vec~\cite{hoang2020cc2vec} attempted to represent code changes more structurally by leveraging ASTs as well. To get the best of both worlds, more recent work tried to combine token information with structure information to obtain a better patch representation~\cite{dong2022fira}.
Finally, several such approaches of patch representation learning have been evaluated on specific tasks, e.g., BATS~\cite{tian2022predicting} for patch correctness assessment and FIRA~\cite{dong2022fira} for patch description generation.
 
On the one hand, token-based approaches for patch representation~\cite{hoang2020cc2vec,xu2019commit,nie2021coregen} lack the rich structural information of source code and intention features inside the sequence is still unexplored.
On the other hand, graph-based representation of patches~\cite{liu2020atom,lin2022context} lacks the context which is better represented by the sequence of tokens~\cite{hoang2020cc2vec,xu2019commit,nie2021coregen} of the patch itself and also the surrounding unchanged code and intention features inside graph changes is also still unexplored.
In conclusion, approaches that try to combine context and AST information to represent patches (\eg FIRA~\cite{dong2022fira}) do not use the intention features of either sequence or graph from the patch but rather rely on representing the code before and after the change while adding some {\em ad-hoc} annotations to highlight the changes for the model.


{\bf This paper.} 
We propose a novel patch representation that tackles the aforementioned problems and provides an extensive evaluation of our approach on three practical and widely used downstream software engineering tasks.
Our approach, \tool, learns to represent patches through a combination of (1) the context around the code change, (2) a novel SeqIntention representation of the sequential patch, and (3) a novel representation of the GraphIntention from the patch.
Our approach enables us to leverage powerful DL models for the sequence intention such as Transformers and similarly powerful graph-based models such as GCN for the graph intention. Additionally, our model is pre-trained and hence task agnostic where it can be fine-tuned for many downstream tasks. 
We provide an extensive evaluation of our model on three popular patch representation tasks: (1) Generating patch description in NL, (2) Predicting patch correctness in program repair, 
and (3) Patch intention detection.

Overall, this paper makes the following contributions: 
\begin{itemize}[leftmargin=*]
        \item [$\blacktriangleright$] {\em A novel representation learning approach for patches}: we combine the context surrounding the patch with a novel sequence intention encoder and a new graph intention encoder to represent the intention of code changes in the patch while enabling the underlying neural models to focus on the code change by representing it explicitly. To that end, we developed: 
        \ding{182}{\em an adapted    Transformer    
        architecture for code sequence intention} to capture sequence intention in patches taking into account not only the changed lines (added and removed) but also the full context (i.e., the  code chunk before the patch application); 
        \ding{183}      {\em an embedding approach for graph intention} to compute embeddings of graph intention  capturing the semantics of code changes.
  
        \item [$\blacktriangleright$] {\em A dataset of parsable patches}: given that existing datasets only provide patches with incomplete details for readily collecting the code before and after the patch, extracting AST diffs was challenging. We therefore developed tool support to enable such collection and produced a dataset of 90k patches, which can be parsed using the Java compiler. 
        \item [$\blacktriangleright$] {\em Extensive evaluation}: we evaluate our approach by assessing its performance on several downstream tasks. For each task, we show how \tool outperforms carefully-selected baselines. We further show that \tool outperforms the state of the art in patch representation learning.
\end{itemize}

\section{\tool}\label{sec:methodology}
\begin{figure}[t]
    \centering
    \includegraphics[width=1\linewidth]{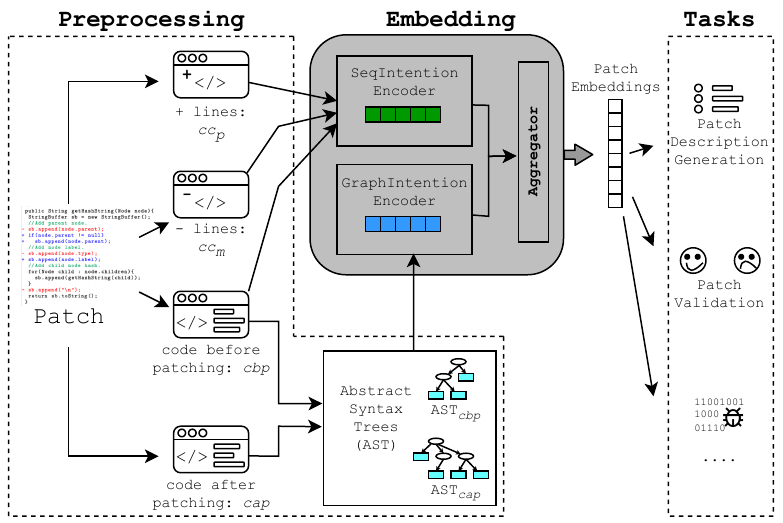}
    \caption{Overview of \tool.}
    \label{fig:pipeline}
\end{figure}

Figure \ref{fig:pipeline} presents the overview of \tool. 
Patches are first preprocessed to split the available information about added ({\tt +}) and removed ({\tt -}) lines, identifying the code context (i.e., the code chunk before applying the patch) and computing the ASTs of the code before and after applying the patches (cf. Section~\ref{sec:preprocess}).
Then, \tool deploys two encoders, which capture sequence intention semantics (cf. Section~\ref{seq-diff}) and graph intention semantics (cf. Section~\ref{sec:graph_extractor}). Those encoded information are aggregated (cf. Section~\ref{sec:assembling}) to produce patch embeddings  that can be applied to various downstream tasks. 
In the rest of this section, we will detail the different components of \tool before discussing the pre-training phase (cf. Section~\ref{sec:pretrain}).






\subsection{Patch Preprocessing} \label{sec:preprocess}
The preprocessing aims to focus on three main information within a patch for learning its representation. 
The code before applying the patch (which provides contextual information of the code change), plus and minus lines (which provide information about the code change operations), and the difference between \textit{AST} graphs before and after patch (which provides information about graph intention in the code). 
Through the following steps we collect the necessary multi-modal inputs (code text, sequence intention, and graph intention) for the learning:

\begin{enumerate}[leftmargin=*]
    \item \textbf{Collect {\tt +/-} lines in the patch.} We scan each patch line. Those starting with a {\tt +} are added to a {\em pluslist}, while those starting with a {\tt -} are added to a {\em minuslist}. Both lists record the line numbers in the patch.  
    \item \textbf{Reconstruct before/after code.} Besides +/- lines, a patch includes unchanged code that are part of the context. We consider that the full context is the code before applying the patch (i.e., unchanged \& minuslist lines). We also construct the code after applying the patch (i.e., unchanged \& pluslist lines). The reconstruction leverages the recorded line numbers for inserting each added/removed line to the proper place and ensure accuracy. 
    \item \textbf{Generate code ASTs before and after patch.} We apply the Javalang~\cite{thunes2013javalang} tool to generate the ASTs for the reconstructed code chunks before and after applying the patch.
    \item \textbf{Construct vocabulary.} Based on the code changes of the patches in the training data, we build a vocabulary using the Byte-Pair-Encoding (\textit{BPE}) algorithm. 
\end{enumerate} 

At the end of this preprocessing phase, for each given patch, we have a set of inputs: \\ $\left \langle cc_p, cc_m, cbp, cap, G_{cbp}, G_{cap}\right\rangle$, where ${cc_p}$ is the sequence of added (+) lines of code, $cc_m$ is the sequence of removed (-) lines of code, $cbp$ is the {\bf c}ode chunk {\bf b}efore the {\bf p}atch is applied, $cap$ is the {\bf c}ode chunk {\bf a}fter the {\bf p}atch is applied, $G_{cbp}$ is the AST graph of $cbp$ and $G_{cap}$ is the graph of $cap$. 

\subsection{Sequence Intention Encoder}
\label{seq-diff}

\begin{figure}[!ht]
    \centering
    \includegraphics[width=0.8\linewidth]{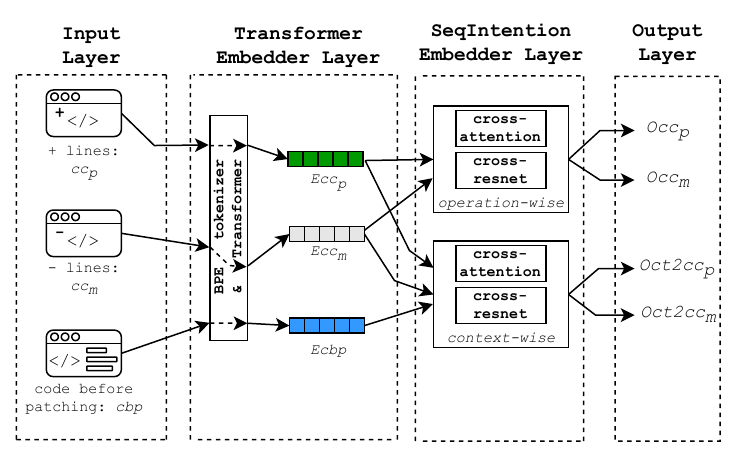}
    \caption{Architecture for the Sequence Intention Encoder.}
    \label{fig:patchformer}
\end{figure}

A first objective of \tool is to build an encoder that is capable of capturing the semantics of the sequence intention in a patch. 
While prior work focuses on {\tt +/-} lines, we postulate that code context is a relevant additional input for better encoding such differences. 
Figure~\ref{fig:patchformer} depicts the architecture of the Sequence intention encoder. 
We leverage the relevant subset of the preprocessed inputs (cf. Section~\ref{sec:preprocess}) to pass to a Transformer embedding layer and further develop a specialized layer, named the {\em SeqIntention embedding layer}, which captures the intention features from the sequence. 

\subsubsection{Input Layer}
The input, for each patch, consists of the triplet $\left \langle cc_m, cc_p, cbp\right\rangle$, where $cc_m$ is the set of removed ({\tt -}) lines, $cc_p$ the set of added ({\tt +})  lines and $cbp$ is the code before patching, which represents the context.

\subsubsection{Transformer Embedding Layer} \label{sec:seq_extractor}
To embed the sequence of code changes, we use a Transformer as the initial embedder. Indeed, Transformers have been designed to capture semantics in long texts and have been demonstrated to be effective for inference tasks~\cite{devlin2018bert,DBLP:conf/dasfaa/WangTZLH22}.

We note that $cc_m \in$ \textit{cbp}.
Assuming that $cc_p$ = $\{token_{p,1}, \dots, token_{p,j}\}$, $cc_m$ = $\{token_{m,1}, \dots, token_{m,k}\}$, \textit{cbp} = $\{token_{cbp,1}, \dots, token_{cbp,l}\}$, where \textit{j, k, l} represent the maximum length of $cc_p$, $cc_m$, and $cbp$ respectively, we use the initial embedding layer the Pytorch implementation of a Transformer to produce first vector representations for each input information as:
\begin{equation}
    E_{X} = Transformer(Init(X;\Theta_1); \Theta_2)
    \label{eq:transformer}
\end{equation}
where $X$ represents an input (either $cc_m$, $cc_p$ or $cbp$); {\it Init} is the initial embedding function; {\it Transformer} is the model based on a transformer architecture; $\Theta_1$ and $\Theta_2$ are the parameters of {\it Init($\cdot$)} and {\it Transformer($\cdot$)}, respectively.

The Transformer embedding layer outputs $E_{cc_p} = [e_{p,1}, e_{p,2}, \dots, e_{p,j}] \in \mathbb{R}^{j\times d_e}$, \\ $E_{cc_m} = [e_{m,1}, e_{m,2}, \dots, e_{m,k}]$ $\in \mathbb{R}^{k\times d_e}$, $E_{cbp} = [e_{cbp,1}, e_{cbp,2}, \dots, e_{cbp,l}]$ $\in \mathbb{R}^{l\times d_e}$, where $d_e$ is the size of the embedding vector. 

\subsubsection{SeqIntention Embedding Layer}
Once the Transformer embedding layer has produced the initial embeddings for the inputs $cc_p$, $cc_m$ and $cbp$, our approach seeks to capture how they relate to each other. Prior works~\cite{shaw2018self, devlin2018bert,xu2019commit,dong2022fira} have proven that self-attention is effective in capturing relationships among embeddings. We thus propose to capture relationships between the added and removed sequences, with the objective of capturing the intention of the code change through the change operations. We also propose to pay attention to context information when capturing the semantics of the sequence intention.

\paragraph{Operation-wise} To obtain the intention of modifications in patches, we apply a cross-attention mechanism between $cc_p$ and $cc_m$. 
To that end, we design a resnet architecture where the model performs residual learning of the importance of inputs (\ie $E_{cc_p}$ and its evolved $\mathcal{v}_p$, which will be introduced below).

To enhance $E_{cc_p}$ into $E_{cc_m}$, we apply a \textbf{cross-attention} mechanism. For the {\it i-th} token  
in $cc_m$, we compute the matrix-vector product, $E_{cc_p} e_{m,i}$, where $e_{m,i} \in \mathbb{R}^{d_e}$ is a vector parameter for {\it i-th} token $\mathcal{i}$ in $cc_m$. We then pass the resulting vector through a softmax operator, obtaining a distribution over locations in the $E_{cc_p}$,

\begin{equation}
    \alpha_\mathcal{i} = SoftMax(E_{cc_p} e_{m,i}) \in \mathbb{R}^{k},
\end{equation}
where SoftMax(\textbf{\em x}) = $\frac{exp(x)}{\Sigma_j exp(x_j)}$. {\em exp(x)} is the element-wise exponentiation of the vector \textit{x}. {\it k} is the length of $cc_m$, The attention $\alpha$ is then used to compute vectors for each token in $cc_m$,
\begin{equation}
    \mathcal{v}_{\mathcal{i}} = \Sigma^{j}_{n=1} \alpha_{\mathcal{i,n}} h_n. 
\end{equation}
where $h_n$ $\in$ $E_{cc_p}$, {\it j} is the length of $cc_p$. In addition, $\mathcal{v}_{\mathcal{i}}$ is the new embedding of {\it i-th} token in $cc_m$ enhanced by semantic of $E_{cc_p}$. 

Then, we get new $cc_m$ embedding $v_m= [\mathcal{v}_1, \dots, \mathcal{v}_k] \in \mathbb{R}^{k \times d_e}$.

Similarly, following steps above, we can obtain new embedding of $cc_p$, $v_p \in \mathbb{R}^{j \times d_e}$, enhanced by the semantic of $cc_m$. 



\begin{figure}[ht]
    \centering
    \includegraphics[width=0.5 \linewidth]{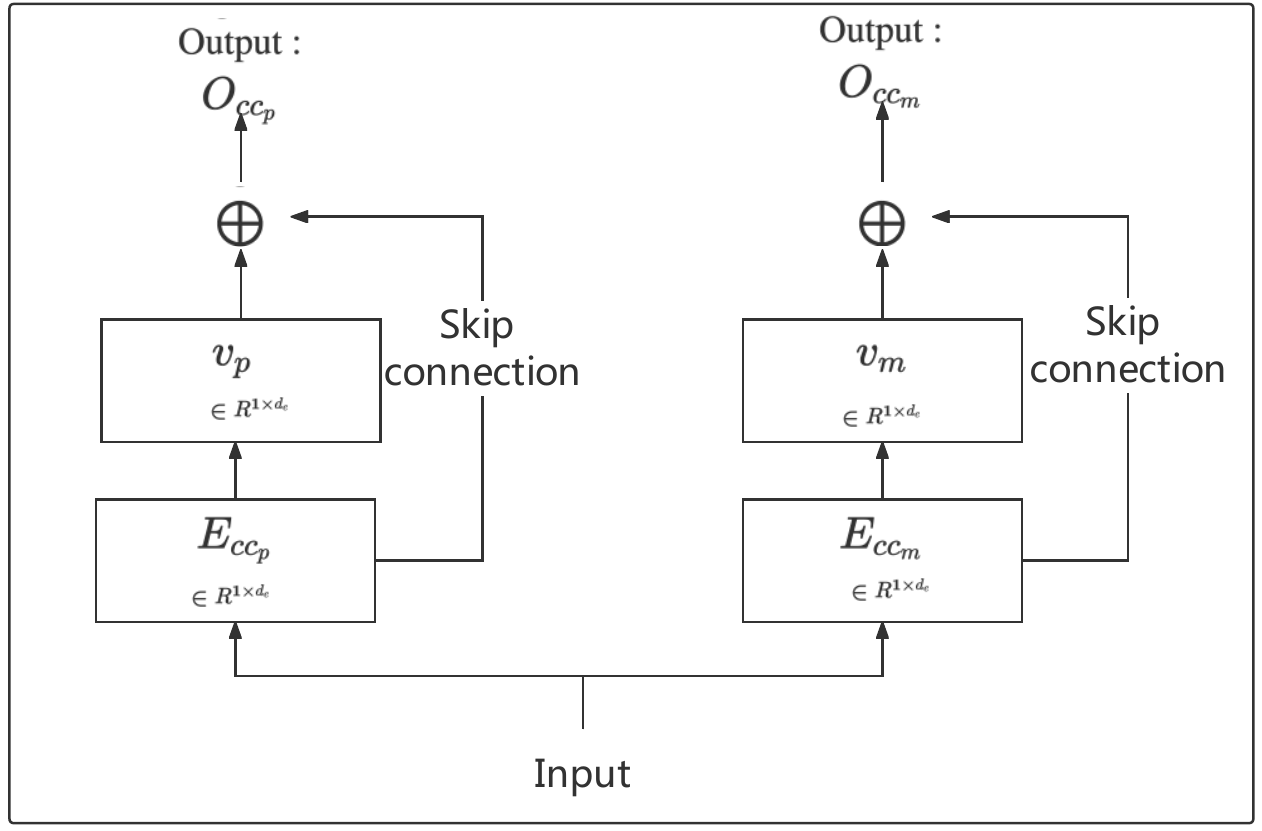}
    \caption{cross-resnet architecture.}
    \label{fig:across-resnet}
\end{figure}

For the combination of $v_p$, $v_m$, $E_{cc_p}$, $E_{cc_m}$,
inspired by \cite{shi2021cast,he2016deep}, we design a \textbf{cross-resnet} for combining $v_p$, $v_m$, $v_{cc_p}$, and $v_{cc_m}$. The pipeline of \textbf{cross-resnet} is shown in Figure \ref{fig:across-resnet}. The process is as follows:
\begin{equation}
\begin{split}
    O_{cc_p}&= f(h(E_{cc_p}) + \mathcal{Add}(E_{cc_p}, v_p))  \\
    O_{cc_m}&= f(h(E_{cc_m}) + \mathcal{Add}(E_{cc_m}, v_m)) 
\end{split}
\label{eq:resnet}
\end{equation}
where {\it h($\cdot$)} is a normalization function in \cite{devlin2018bert}; $\mathcal{Add}(\cdot)$ is the adding function, {\it f($\cdot$)} represents \textit{RELU} \cite{glorot2011deep} activation function.

Finally, we obtain output $O_{cc_m}$ and $O_{cc_p}$. 

\paragraph{Context-Wise} Similar to operation-wise block, we enhance the contextual information into modified lines by \textbf{cross-attention} and \textbf{cross-resnet} blocks. The computation process is as follows:
\begin{equation}
    \begin{split}
        O_{ct2cc_p} &= f(h(E_{cc_p}) + \mathcal{Add}(E_{cc_p}, E_{cbp}))  \\
        O_{ct2cc_m} &= f(h(E_{cc_m}) + \mathcal{Add}(E_{cc_m}, E_{cbp}))
    \end{split}
    \label{eq:context-wise}
\end{equation}

where $E_{cbp}$ is the embedding of {\it cbp} calculated by Equation \ref{eq:transformer}; $O_{ct2cc_p}$ represents the vector of context-enhanced plus embedding and $O_{ct2cc_m}$ is the vector of context-enhanced minus embedding.



\subsection{Graph Intention Encoder} \label{sec:graph_extractor}
\begin{figure*}[ht]
    \centering
    \includegraphics[width=0.75\linewidth]{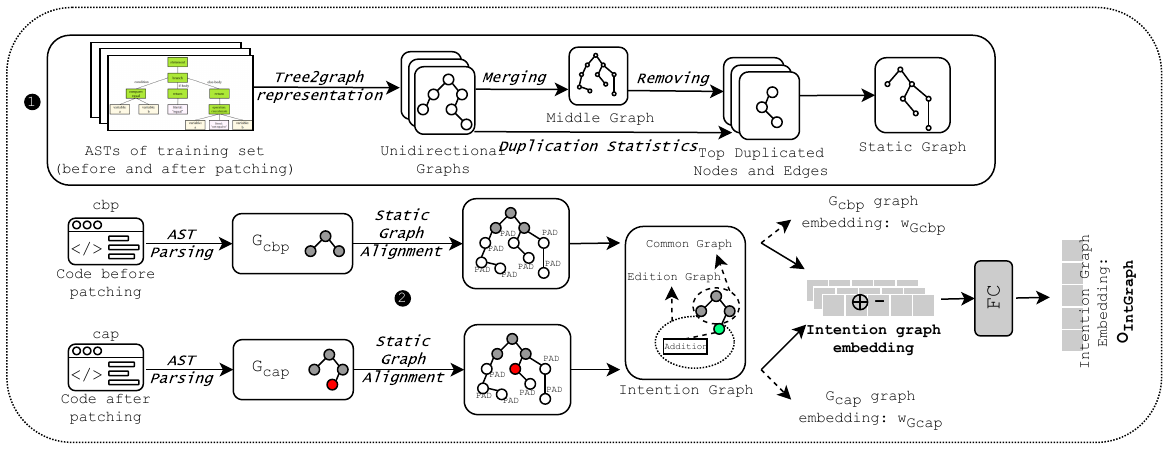}
    \caption{Architecture for the Graph Intention Encoder.}
    \label{fig:astdiff}
\end{figure*}

Concurrently to encoding sequence information from code changes, we propose to also capture the graph intention features of the structural changes in the code when the patch is applied. To that end, we rely on a Graph Convolutional Network (GCN) architecture, which is widely used to capture dynamics in social networks, and is effective for typical graph-related tasks such as classification or knowledge injection~\cite{kipf2016semi,zhang2019graph,zhang2019ernie}. 
Once the GCN encodes the graph nodes, the produced embeddings can be used to assess their relationship via computing their cosine similarity scores~\cite{luo2018cosine}. 
Concretely, in \tool, we use a GCN-based model to capture the graph intention features. The embedder model was trained by inputting a static graph, a graph resulting from the merge of all sub-graphs from the training set. 
Overall, we implement this encoding phase in two steps: building the static graph, performing graph learning and encoding the graph intention (cf. Figure~\ref{fig:astdiff}).

\subsubsection{Graph Building}
To start, we consider the $G_{cbp}$ and $G_{cap}$ trees, which represent in graph forms.

\begin{figure}[!ht]
    \centering
    \includegraphics[width=0.9\linewidth]{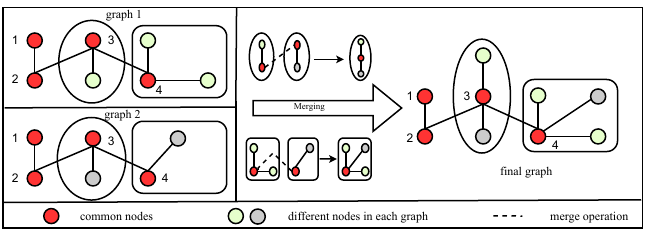}
    \caption{An example of merging graphs.}
    \label{fig:merge}
\end{figure}

\noindent
\ding{182} \textbf{Static Graph building:} 
Each patch in the dataset can be associated to two graphs: $G_{cbp}$ and $G_{cap}$, which are obtained by parsing the $cbp$ and $cap$ code snippets.
After collecting all graphs (which are unidirectional graphs) for the whole training set, we merge them into a ``big" graph by iteratively linking the common nodes. In this big graph, each distinct code snippet AST-inferred graph is placed as a distinct sub-graph. Then, we will merge the graphs shown in Figure~\ref{fig:merge} which illustrates the merging progress of two graphs: if a node $\mathcal{N}$ has the same value, position, and neighbors in both ASTs, it will be merged into one (e.g., red nodes 1 and 2). However, when a common node has different neighbors between the ASTs (e.g., red nodes 3 and 4), the merge keeps one instance of the common node but includes all neighbors connected to the merged red nodes (\ie all green and grey nodes are now connected to red nodes 3 and 4, respectively). After iterating over all graphs, we eventually build the static graph.

However, on the one hand, some nodes in most subgraphs such as `prefix\_operators', `returnStatement', and `StatementExpression' are not related to the semantics of the patch. On the other hand, as data statistics in our study, 97.2\% nodes in the initial graphs (ASTs) extracted from code are from parser tools instead of patches, which means these nodes are rarely related with the semantic of the patch. Thus, as shown in step 1 in Figure~\ref{fig:astdiff}, we remove nodes whose children do not contain words in the patch to reduce the size of the graph because these nodes will be considered noise in our research.

In the remainder of this paper, we refer to the final graph as the \textit{static graph} $\mathcal{G}$=$(\mathcal{V}, \mathcal{E})$, where $\mathcal{V}$ is the set of nodes and $\mathcal{E}$ is the set of edges. 

\noindent
\ding{183} \textbf{Graph Alignment to the Static Graph:} 
GCN requires that the input graphs are all of the same size~\cite{kipf2016semi}. Yet, the graphs built using the graphs of $cbp$ and $cap$ do not have as many nodes and edges as the static graph used for training the GCN network.
Consequently, we propose to use the static graph to initialize all graphs. 
Given the global static graph $\mathcal{G}_{global} = (\mathcal{V}_g, \mathcal{E}_g)$ and an AST graph $\mathcal{G}_{local} =(\mathcal{V}_l, \mathcal{E}_l)$, we leverage the {\it VF} graph matching algorithm~\cite{cordella1999performance} to find the most similar sub-graph with $\mathcal{G}_{local}$ in $\mathcal{G}_{global}$:
\begin{equation}
    subGraph = VFG(\mathcal{G}_{local}, \mathcal{G}_{global}).
    \label{eq:subgraphmatching}
\end{equation}
where $VFG(\cdot)$ is the function representing the VF matching algorithm~\cite{networkx}.
 The matched sub-graph is a subset of both  $\mathcal{G}_{local}$ and $\mathcal{G}_{global}$: some nodes of  $subGraph$ will be in $ \mathcal{G}_{local}$ but not in $\mathcal{G}_{global}$. 
We then align $\mathcal{G}_{local}$ to the same size of $\mathcal{G}_{global}$ as follows: we use the [PAD] element to pad the node of subGraph to the same size of $\mathcal{G}_{global}$ and then we obtain $\mathcal{G}_{l_{PAD}}$. 
Therefore, $\mathcal{G}_{l_{PAD}}$ keeps the same size and structure of the static graph $\mathcal{G}_{global}$. 
Eventually, all graphs are aligned to the same size of $\mathcal{G}_{global}$ and the approach can meet the requirements for GCN computation for graph learning.

\subsubsection{Graph Learning}
Inspired by \cite{zhang2019graph}, we build a deep graph convolutional network based on the undirected graph formed following the above construction steps to further encode the contextual dependencies in the graph. Specifically, for a given  undirected graph $\mathcal{G}_{l_{PAD}}$ = ($\mathcal{V}_{l_{PAD}}$, $\mathcal{E}_{l_{PAD}}$), 
let $\mathcal{P}$ be the renormalized graph laplacian matrix \cite{kipf2016semi} of $\mathcal{G}_{l_{PAD}}$:
\begin{equation}
    \begin{aligned}
        \mathcal{P} &= \hat{\mathcal{D}}^{-1/2}\hat{\mathcal{A}}\hat{\mathcal{D}}^{-1/2}  \\
        &=(\mathcal{D}+\mathcal{L})^{-1/2}(\mathcal{A}+\mathcal{L})(\mathcal{D}+\mathcal{L})^{-1/2} 
    \end{aligned}
\end{equation}
where $\mathcal{A}$ denotes the adjacency matrix, $\mathcal{D}$ denotes the diagonal degree matix of the graph $\mathcal{G}_{l_{PAD}}$, and $\mathcal{L}$ denotes the identity matrix. The iteration of GCN through its different layers is formulated as:
\begin{equation}
    \mathcal{H}^{(l+1)} = \sigma(((1-\alpha)\mathcal{P}\mathcal{H}^{(l)} + \alpha\mathcal{H}^{(0)})((1-\beta^{(l)})\mathcal{L} + \beta^{(l)}\mathcal{W}^{(l)}))
\end{equation}
where $\alpha$ and $\beta^{(l)}$ are two hyper parameters, $\sigma$ denotes the activation function and $\mathcal{W^{(l)}}$ is a learnable weight matrix. Following GCN learning, we use the average embedding of the graph to represent the semantic of structural information in code snippet:
\begin{equation}
    w_{G} = \frac{1}{L}\Sigma_{i=1}^L(\mathcal{H_i}).
    \label{eq:astgcn}
\end{equation}

Thus, at the end of the graph embedding, we obtain representations for $G_{cbp}$ and $G_{cap}$,
\ie $w_{G_{cbp}}$ and $w_{G_{cap}}$ $\in \mathbf{R}^{1\times d_e}$.

\begin{figure}[!h]
    \centering
    \includegraphics[width=0.5 \linewidth]{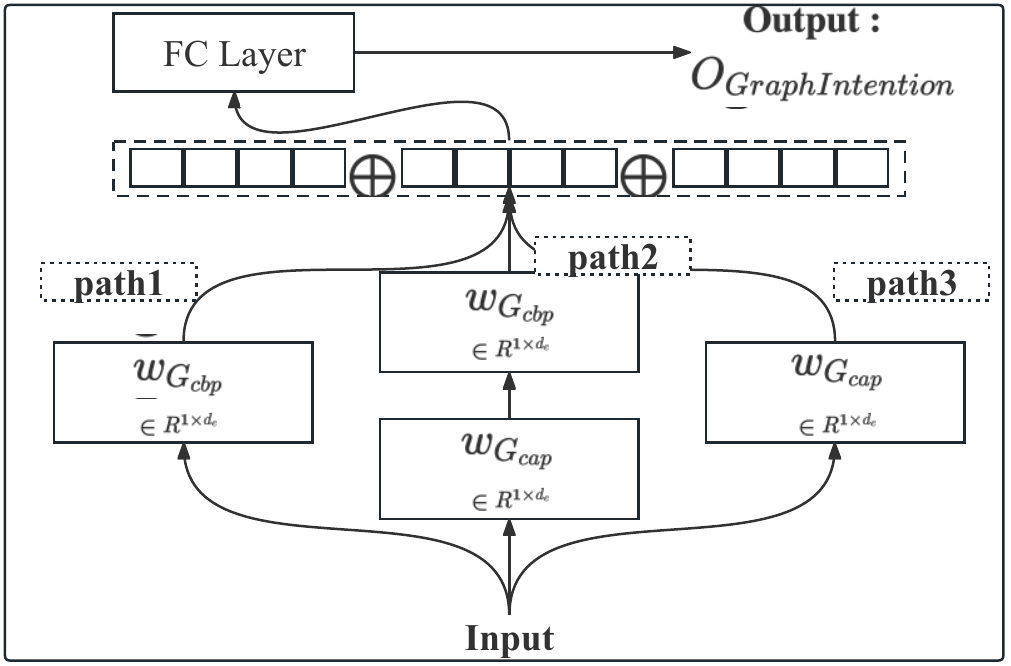}
    \caption{graph-cross-resnet architecture.}
    \label{fig:graph-across-resnet}
\end{figure}

\subsubsection{Graph Intention Encoding}
Once we have computed the embeddings of the code snippets before and after patching, (i.e., the embeddings of $cbp$ and $cap$), we must get the representation of their differences to encode the intention inside the graph changes. To that end, similarly to the previous cross-resnet for sequence intention, we design a \textbf{graph-cross-resnet} operator which ensembles the semantic of $w_{G_{cbp}}$ and $w_{G_{cap}}$. Figure~\ref{fig:graph-across-resnet} illustrates this crossing. In this graph-cross-resnet, the model can choose and highlight a path automatically by the backpropagation mechanism. The {\em GraphIntention} is therefore calculated as follows:
\begin{equation}
    \begin{split}
        path1 &= w_{G_{cbp}} \\
        path2 &= \mathcal{Add}(w_{G_{cbp}},w_{G_{cap}}) \\
        path3 &= w_{G_{cap}} \\
        O_{GraphIntention} &= \mathcal{FC}(f(\mathcal{Add}(path1, path2, path3)),\Theta_3) \\
    \end{split}
\end{equation}
where $\mathcal{FC}$ is a fully-connected layer and $\Theta_3$ is the parameter of $\mathcal{FC}$.

At the end, the {graph-cross-resnet} component outputs the sought graph intention embedding: {\em GraphIntention}.

\subsection{Aggregating Multimodal Input Embeddings} 
\label{sec:assembling}
With the sequence intention encoder and the graph intention encoder, we can produce for each patch several embeddings of different input modalities (code sequences and graphs) that must be aggregated into a single representation. 

Concretely, we use the $\mathcal{Add}$ aggregation function to merge the {\em SeqIntention} embeddings (combination of $O_{cc_p}$, $O_{cc_m}$, $O_{ct2cc_p}$ and  $O_{ct2cc_m}$ - cf. Equations~\ref{eq:resnet} and \ref{eq:context-wise}) and {\em GraphIntention} embedding $O_{GraphIntention}$ before outputing the final representation $E_{\tool}$. 
Actually, we use $E_{\tool}$ as the representation of patch out of the model.

\subsection{Pre-training}\label{sec:pretrain} 
\tool is an approach that is agnostic of the downstream tasks. We propose to build a pre-trained model with the collected dataset of patches. The objective is to make the model learn contextual semantics of code, which can help improve the efficiency of representing patches. 

The pre-training task is masked token prediction. Indeed, a popular bidirectional objective function is driven by the masked language model (MLM; \cite{devlin2018bert}), which aims to predict masked word based on its surrounding context, i.e., compute the probability $P(x_i|x_1, \dots, x_{i-1}, x_{i+1},\dots, x_n)$.

Inspired by previous prior works in model pre-training~\cite{feng2020codebert,elnaggar2021codetrans,devlin2018bert,zhang-etal-2019-ernie}, we only employ predicting [MASK]s as our pre-training strategy. 
We randomly mask some words in the source code. The target is then to use the contextual information to predict the masked word as accurately as possible.


We take the pre-training as a translation task and employ the popular encoder-decoder architecture as our pipeline. We use our \tool architecture as our encoder and the BERT Transformer model~\cite{devlin2018bert} as the decoder, where they share the same parameters. Apart from that, encoder and decoder share vocab. Finally, we start the decoding process with an initial \texttt{<s>} token to generate the code sequence word by word:
\begin{equation}
    index = argmax(p(y_t|y_{t-1},\cdots, y_1, E_{\tool}))
    \label{eq:pretrain}
\end{equation}
where $y_i (i=1,\cdots,t)$ is a one-dimension distribution where the distribution size equals the length of vocab. The index is the number of the element which is the max one in $y_i$. Then we can return the corresponding word with the same index in vocab. Furthermore, we employ the Cross-Entropy \cite{rubinstein1999cross} as the loss function and apply Adam algorithm \cite{kingma2014adam} as the optimizer.



\subsection{Fine-tuning for Different Tasks} \label{sec:finetune}
\tool works as an embedder for patches. It is pre-trained by unsupervised learning contextual semantics with collected datasets of patches. Thus, given a patch, \tool is able to generate its task-agnostic embedding.

We can now use the generated embeddings for various downstream tasks. We now describe how patch representations produced by \tool are adopted to address the three popular patch-related downstream tasks that we investigate in this work: patch description generation \cite{dong2022fira,xu2019commit,hoang2020cc2vec,nie2021coregen}, patch correctness assessment \cite{tian2022change}, Just In Time (JIT) defect prediction \cite{hoang2019deepjit, hoang2020cc2vec}.

\subsubsection{Patch Description Generation}
This task aims to generate a natural-language description for a given patch. The patch description generation task is highly similar to the pre-training task (cf. \ref{sec:pretrain}). The training dataset of this task is bimodal. It includes patches associated to descriptions. 
We input the patch and fine-tune the model to generate pseudo descriptions that are as much as possible similar to the ground-truth descriptions. The  fine-tuning process \tool is therefore formulated as for pre-training masked word prediction of Equation \ref{eq:pretrain}.

\subsubsection{Patch Correctness Assessment}
This task is relevant in the program repair community where generated patches must be predicted as correct or not for a given bug. 
It is a binary classification task. 
Given the patch embedding produced by \tool and a bug report embedding produced by BERT~\cite{devlin2018bert}, we concatenate them and apply a fully-connected layer to classify it as correct or not:
\begin{equation}
    \hat{y_i} = sigmoid(E_{patch_i}\oplus E_{bugReport_i})
    \label{eq:classify}
\end{equation}
where $E_{patch_i}$ is the \tool's embedding of patch $i$, \linebreak $E_{bugReport_i}$ is the embedding of the associated bug report, and $\oplus$ is the concatenation operator.

To optimize our model for the task of patch correctness assessment, we opt for a categorical cross-entropy loss function as follows:
\begin{equation}
    \mathcal{L}_a = -\mathop{\Sigma}\limits_{i=1}^n (y_i \times (\ln(\hat{y}_i)) + (1-y_i) \times \ln(1-\hat{y}_i))
    \label{eq:lossclassify}
\end{equation}
where $\mathcal{L}_a$ is loss value, {\it n} is the size of the dataset, $y_i$ is the ground-truth label, $\hat{y}_i$ is the predicted label.

\section{Experimental Design}\label{sec:exp_design}
We provide the implementation details (cf. Sec.~\ref{sec:implementation}), discuss the research questions (cf. Sec.~\ref{sec:rqs}), and present the baselines (cf.~Sec.~\ref{sec:baselines}), the datasets (cf.~Sec.~\ref{sec:datasets}), and the metrics (cf.~Sec.~\ref{sec:metrics}).

\subsection{Implementation}
\label{sec:implementation}
In the pre-training phase used for the Sequence Intention Embedding step, we apply a beam search \cite{vijayakumar2016diverse} for the best performance in predicting the masked words. The beam size was set to 3. The dimension of the hidden layer output in models is set to 512, and the default value of dropout rate is set to 0.1. For the Transformer, we apply 6 heads for the multi-header attention module and 4 layers for the attention.

For the Graph Intention Embedding step, we use javalang~\cite{thunes2013javalang} to parse code fragments and collect ASTs. We build on graph manipulation packages (\ie networkx~\cite{networkx}, and dgl~\cite{dgl}) to represent these ASTs into graphs. 

\tool's training involves the Adam optimizer~\cite{kingma2014adam} with learning rate 0.001. All model parameters are initialized using Xavier algorithm~\cite{glorot2010understanding}. 
All experiments are performed on a server with Intel(R) Xeon(R) CPU E5-2698 v4 @ 2.20GHz, 256GB physical memory, and one NVIDIA Tesla V100 GPU with 32GB memory.


\vspace{-2mm}
\subsection{Research questions}
\label{sec:rqs}

{\bf RQ-1}: {\em How effective is {\em \tool} in learning patch representations?} 
{\bf RQ-2}: {\em What is the impact of the key design choices on the performance of {\em \tool}? } \\
{\bf RQ-3}: {\em To what extent is {\em \tool} effective on independent datasets?} 

\vspace{-2mm}
\subsection{Baselines}
\label{sec:baselines}




We consider several SOTA models as baselines. 
We targeted approaches that were specifically designed for patch representation learning (e.g., CC2Vec) as well as generic techniques (e.g., NMT) that were already applied to patch-related downstream tasks. We finally consider  recent SOTA for patch-representation approaches  (e.g., FIRA) for specific downstream tasks. 
\vspace{-1mm}
\begin{itemize}[leftmargin=*]
    \item \textbf{NMT} technique has been leveraged by Jiang~\etal~\cite{jiang2017automatically} for translating code commits into commit messages.
    \item \textbf{NNGen}~\cite{liu2018neural} is an IR-based commit message prediction technique. 
    \item \textbf{CoDiSum}~\cite{xu2019commit} is an encoder-decoder based model with multi-layer bidirectional GRU and copying mechanism \cite{see2017get}.
    \item \textbf{CC2Vec}~\cite{hoang2020cc2vec} learns a representation of code changes guided by commit messages. It is the incubent state of the art that we aim to outperform on all tasks.
    \item \textbf{CoRec}~\cite{wang2021context} is a retrieval-based context-aware encoder-decoder model for commit message generation.
    
    \item \textbf{Coregen}~\cite{nie2021coregen} is a \textbf{pure Transformer-based} approach for representation learning targeting commit message generation. 
    \item \textbf{FIRA}~\cite{dong2022fira} is a graph-based code change representation learning approach for commit message generation.
    \item \textbf{BERT}~\cite{devlin2018bert} is a state of the art unsupervised learning based Transformer model widely used for text processing.
    \item \textbf{ATOM}~\cite{liu2020atom} is a commit message generation techniques, which builds on abstract syntax tree and hybrid ranking.
    \item \textbf{CCRep}~\cite{liu2023ccrep} is an innovative approach that uses pre-trained models to encode code changes into feature vectors, enhancing performance in tasks like commit message generation, etc.
    
\end{itemize}

\vspace{-2mm}

\subsection{Datasets} \label{sec:dataset}
\label{sec:datasets}
\noindent
\textbf{Patch description generation:}
We build on prior benchmarks~\cite{dyer2013boa, hoang2020cc2vec, liu2018neural, dong2022fira} by focusing on Java samples and reconstructing snippets to make them parsable for AST collection. Eventually, our dataset includes 90,661 patches and their associated descriptions. 


\noindent
\textbf{Patch correctness assessment:}
We leverage the largest dataset in the literature to date, which includes deduplicated 11,352 patches (9,092 Incorrect and 2,260 Correct) released by Tian~{\em et al.}~\cite{tian2022change}.

\vspace{-2mm}
\subsection{Metrics}
\label{sec:metrics}

\noindent
\textbf{ROUGE}~\cite{rouge2004package} is one set of metrics for comparing automatic generated text against the reference (human-produced) ones. We focus on ROUGE-L which computes the Longest Common Subsequence.

\noindent
\textbf{BLEU}~\cite{papineni2002bleu} is a classical metric to evaluate the quality of machine translations. It measures how many word sequences from the reference text occur in the generated text and uses a (slightly modified) n-gram precision to generate a score. 

\noindent
\textbf{METEOR}~\cite{banerjee2005meteor} is an F-Score-Oriented metric for measuring the performance of text-generation models.



\noindent
{\bf +Recall} and {\bf -Recall}~\cite{tian2022change} are specific metrics for patch correctness assessment. +Recall ({-Recall}) measures to what extent correct (incorrect) patches can be predicted (filtered out).
    

\vspace{-1mm}
\section{Experimental Results}\label{sec:evaluation}

\subsection{[RQ-1]: Performance of \tool 
}
[\textbf{Experiment Goal}]: 
We assess the effectiveness of the embeddings learned by \tool on three popular and widely used software engineering tasks: (RQ-1.1) Patch description generation , (RQ-1.2) Patch correctness assessment, and (RQ-1.3) Patch intention detection.
We compare \tool against the relevant SOTA.



\noindent
\textbf{[Experiment Design (RQ-1.1)]:}
We employ the dataset from FIRA. 
As Xu et al.~\cite{dong2022fira} have previously assessed FIRA and other baseline methods using this dataset, we directly reference the evaluation results of all the baselines from Table IV of the FIRA paper. The dataset contains 75K, 8K and 7.6K commit-message pairs in the training, validation and
test sets, respectively.
We evaluate the generated patch descriptions in the test set using the {BLEU}, {ROUGE-L}, and {METEOR} metrics. 

Note that we distinguish between baseline generation-based methods and retrieval-based ones. In generation-based baselines, a patch description is actually synthesized, while in retrieval-based baselines, the approach selects a description text from an existing corpus (e.g., in the training set). For fairness, we build two distinct methods using \tool's embeddings. The first method is generative and follows the fine-tuning process described in Section~\ref{sec:finetune}. The second method is an IR-based approach, where, following the prior work~\cite{hoang2020cc2vec}, we use \tool as the initial embedding tool and implement a retrieval-based approach to identifying a relevant description in the training set: the description associated with the training set patch that has the highest similarity score with the test set patch is outputted as the "retrieved" description. 

\begin{figure}[!h]  
	\centering
	\subfigure[Generation]{
		\includegraphics[width=4.0cm]{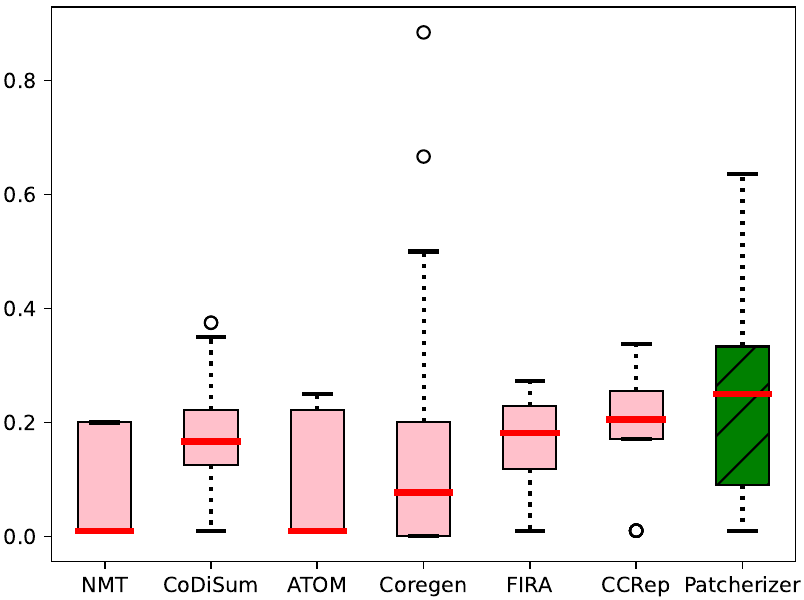}
		\label{fig:gen}
	}
	\subfigure[Retrieval]{
		\includegraphics[width=4.0cm]{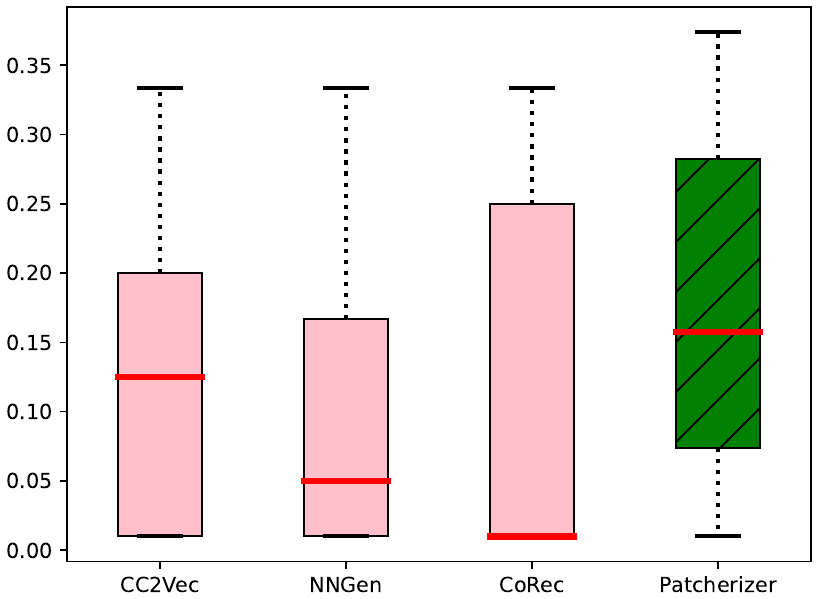}
		\label{fig:ret}
	}
	\caption{Comparison of the distributions of BLEU scores for different approaches in patch description generation} 
	\label{fig:1.2}	
\end{figure}

\noindent
\textbf{[Experiment Results (RQ-1.1)]:}
Table~\ref{tab:T1D2} presents the average scores of the different metrics with the descriptions generated by \tool and the relevant baselines. 
\tool outperforms all the compared techniques on all metrics, with the exception of FIRA on the ROUGE-L metric. 
The superior performance of \tool on generation-based and retrieval-based methods, as illustrated by the distribution of BLEU scores in Figure~\ref{fig:1.2},  further suggest that the produced embeddings are indeed effective. 

\begin{table}[!ht]
\scriptsize
    \centering
    \caption{Performance Results of patch description generation.}
    \resizebox{0.9\linewidth}{!}{
    \begin{tabular}{llrrr}
    \toprule
    Type  & Approach   & Rouge-L  (\%) &  BLEU (\%) & METEOR (\%)\\
    \midrule
    \multirow{7}{*}{\rotatebox{90}{Generation}}
    &  NMT~\cite{jiang2017automatically}  & 7.35 & 8.01 & 7.93\\
    &  Codisum~\cite{xu2019commit}  & 19.73 & 16.55 & 12.83\\ 
    &  ATOM~\cite{liu2020atom}   & 10.17  & 8.35 & 8.73 \\
    &  FIRA~\cite{dong2022fira}   & 21.58 & 17.67 & 14.93\\  
    &  CoreGen~\cite{nie2021coregen} (Transformer) & 18.22 & 14.15  & 12.90\\
    &  CCRep~\cite{liu2023ccrep}  & 23.41  & 19.70   & 15.84\\
    \cmidrule{2-5}
    &  \tool & \cellcolor{black!25} 25.45  & \cellcolor{black!25} 23.52 &  \cellcolor{black!25} 21.23\\
      \midrule
    \multirow{4}{*}{\rotatebox{90}{Retrieval}}
      & CC2Vec~\cite{hoang2020cc2vec}   & 12.21 &12.25   &11.21 \\
      &  NNGen~\cite{liu2018neural}   & 9.16 &9.53 & 16.56\\
      &  CoRec~\cite{wang2021context}   & 15.47 &13.03 & 12.04\\\cmidrule{2-5}
      & \tool  & \cellcolor{black!25}17.32   &\cellcolor{black!25} 15.21 & \cellcolor{black!25}17.25\\
    \bottomrule
    \end{tabular}}
    \begin{tablenotes}
    \footnotesize
    \item ``\textbf{Generation}" for generation-based strategy. Given fragments of codes, ``Generation" methods generate messages from scratch.
    \item ``\textbf{Retrieval}" for retrieval-based approaches. Given fragments of codes, ``\textbf{Retrieval}" approaches return the most similar message from the training dataset.
    
    \end{tablenotes}
    \label{tab:T1D2}
\end{table}

\begin{figure}[!t]
    \centering
\begin{lstlisting}[language=diff,linewidth={\linewidth},frame=tb]
--- s0.java 2022-08-20 17:00:26.000000000 +0200
+++ t0.java 2022-08-20 17:01:04.000000000 +0200
@@ -1,6 +1,6 @@
    public abstract class HardwareRenderer {
        public static boolean sSystemRendererDisabled = false ; 
-       public static boolean sUseRenderThread = true ; 
+       public static boolean sUseRenderThread = false ; 
        private boolean mEnabled ; 
        private boolean mRequested = true ; 
    }
\end{lstlisting}
    \includegraphics[width=0.7\linewidth]{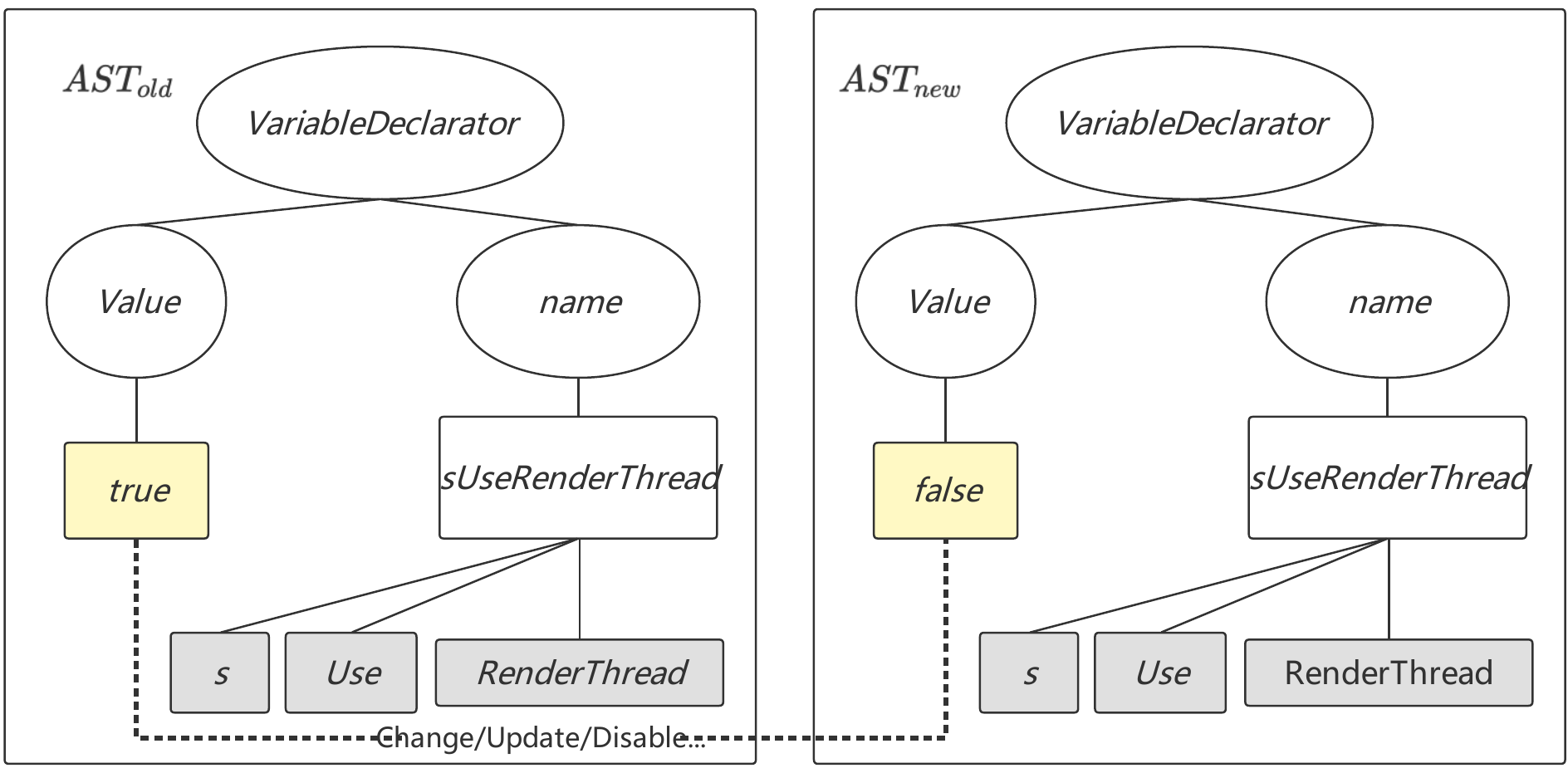}

\vspace{.2em}
 \footnotesize
 \begin{tabular}{ll}
   \toprule 
  Source & Patch description\\\midrule
  \textbf{Ground truth} & \textbf{Disable RenderThread} \\
    CC2Vec & Fix test data so that it can be compiled \\
    FIRA & fix the in \\
    CCRep & update suserenderthread \\
    \tool & disable renderthread \\
    \bottomrule
    \end{tabular}
    \caption{Illustrative example of patch description generation}
    \label{fig:java0}
\end{figure}

In Figure~\ref{fig:java0}, we provide an example result of generated description by \tool, by the CC2Vec strong baseline (using retrieval-based method) and by the FIRA and CCRep state-of-the-art approach 
 (using generation-based method) for patch description generation.
 \tool succeeds in actually generating the exact description as the ground truth commit message, after taking into account both sequential and structural information. 
 By observing the graph intention and sequence intention, we can see that the model found that the only change is that the node {\tt true} has been changed/updated/disabled to {\tt false}. 
 Finally, the sequence intention embedding would make \tool recognize that the carrier of {\tt true} and {\tt false} is {\tt RenderThread} based on BPE splitting.

\find{{\bf \ding{45} Answer to RQ-1.1: }$\blacktriangleright$ 
\tool's embeddings are effective for patch description generation yielding the best scores for BLEU, ROUGE-L, and METEOR metrics.$\blacktriangleleft$ }

\noindent

\noindent
\textbf{[Experiment design (RQ-1.2)]:}
Tian et al.~\cite{tian2020evaluating} proposed to leverage the representation learning (embeddings) of the patches to assess patch correctness. Following up on their study, we use the patch embeddings produced by CC2Vec, BERT, CCRep, and \tool (cf. Section~\ref{sec:finetune}) to train three classifiers to classify APR-generated patches as correct or not and we experiment with two supervised learning algorithms: Logistic regression (LR) and XGBoost (XGB).
To perform a realistic evaluation, we split the patches dataset by bug-id into 10 groups to perform a 10-fold-cross-validation experiment similar to previous work~\cite{tian2022change}. 
In this splitting strategy, all patches for the same bug are either placed in the training set or the testing set to ensure that there is no data leakage between the training and testing data.
We then measure the performance of the classifiers using {+Recall}, {-Recall}, {AUC}, and {F1}.



    

\noindent
\textbf{[Experiment Results (RQ-1.2)]:}
Table~\ref{tab:patch_correctness} shows the results of this experiment.
Both classifiers, LR and XGB, when trained with \tool embeddings largely outperform the classifiers that are trained with BERT or CC2Vec embeddings, which achieved SOTA results in literature~\cite{tian2020evaluating}.

\begin{table}[!h]
	\centering
	\caption{Performance of Patch Correctness Classification}
	\label{tab:patch_correctness}
	\resizebox{0.6\linewidth}{!}
	{
	\begin{threeparttable}
		\begin{tabular}{llrrrr}
			\toprule
            Classifier & Model & AUC & F1 & +Recall & -Recall \\
			\midrule
			\multirow{4}{*}{LR} & CC2Vec & 0.75 & 0.49 & 0.47 & 0.85  \\
			 & BERT  & 0.83 & 0.58 & 0.81 & 0.65  \\
              & CCRep  & 0.86 & 0.67 & 0.74 & 0.83  \\
            & \tool  & \cellcolor{black!25}0.96 & \cellcolor{black!25}0.82 & \cellcolor{black!25}0.87 & \cellcolor{black!25}0.91  \\
			\midrule
            \multirow{4}{*}{XGB} & CC2Vec & 0.81 & 0.55 & 0.50 & 0.89  \\
			 & BERT & 0.84 & 0.61 & 0.64 & 0.85  \\
              & CCRep & 0.82 & 0.63 & 0.59 & 0.88   \\
            & \tool  & \cellcolor{black!25}0.90 & \cellcolor{black!25}0.67 & \cellcolor{black!25}0.66 & \cellcolor{black!25}0.90  \\
		   \bottomrule
		\end{tabular}
	\end{threeparttable}
	}
 \vspace{2mm}
\end{table}

\noindent


\find{{\bf \ding{45} Answer to RQ-1.2: }$\blacktriangleright$ 
Patch embeddings generated by \tool achieve SOTA results in the task of patch-correctness assessment, largely outperforming SOTA embedding models.
$\blacktriangleleft$ }
\noindent
\textbf{[Experiment Goal (RQ-1.3)]:} Previous work introduces that the patch has its intention and detecting the intention of the patch can help the model understand the semantics of the patch (\ie template-based works \cite{buse2010automatically,cortes2014automatically} and generation-based works \cite{dong2022fira,xu2019commit}). Thus, efficiency of patch intention detection can be used to measure if the patch representation model is good or not.  

\noindent
\textbf{[Experiment Results (RQ-1.3)]:}

We scan all words across three datasets in our work and figure out that patches are mainly related to four types: {\tt fix, remove, add}, and {\tt update}. However, {\tt fix} is highly related to all other three frequent words, because {\tt fix} can be used to update, remove or add. Therefore, we select {\tt add, remove, update} as our main detected intentions. In this section, we aim to explore how \tool performs against the representative models {\tt CC2Vec} and {\tt CCRep} on distinguishing the intention of patches.

We trained the three models (i.e., \tool, {\tt CC2Vec}, and {\tt CCRep}) on a large dataset proposed in~\cite{dong2022fira}.
Then, we assess the patch intention detection ability of these models on the CC2Vec dataset~\cite{hoang2020cc2vec}.
We find  that  572 patches contain {\tt add},  {\tt remove}, or {\tt update} keywords (\ie 201 for {\tt add}, 341 for {\tt remove}, 30 for {\tt update}). 
Then, we use the three models to embed these 572 patches and obtain corresponding high-dimensional vectors. 
We employ {\tt t-SNE}~\cite{van2008visualizing} to reduce the dimensionality for better visualization.  

\begin{figure*}[!t]
     \includegraphics[scale=0.47]{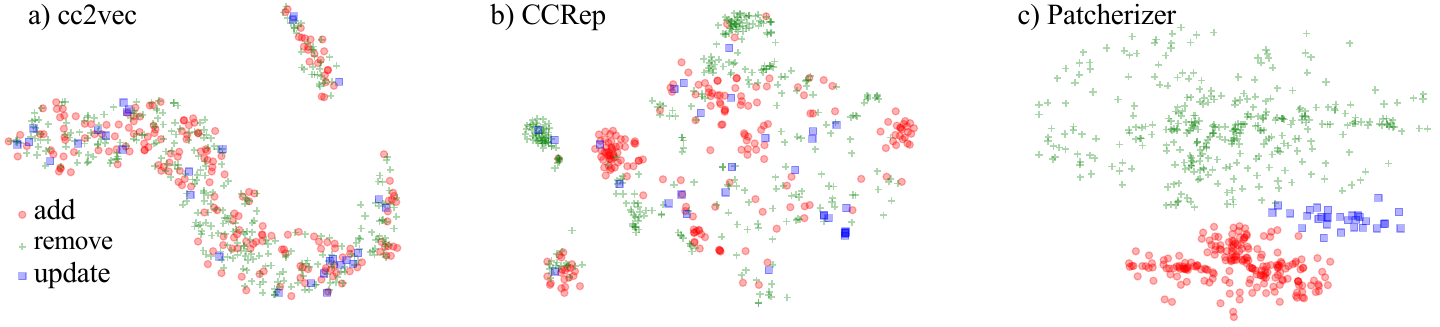}
     \caption{Visualization of Patch intention recognition by different models.}
    \label{fig:intentiondetection}
 \end{figure*}

Figure~\ref{fig:intentiondetection} shows the {\tt t-SNE} visualized results of {\tt CC2Vec}, {\tt CCRep} and \tool.
The red color represents {\tt add} function, the green color represents {\tt remove} function, and the blue color represents {\tt update} function. We see that \tool separates {\tt add} and {\tt remove} better than {\tt CC2Vec} and {\tt CCRep}. Furthermore, both {\tt CC2Vec} and {\tt CCRep} fail to separate {\tt update} from the other two functions. The reason may be that {\tt update} functions can be {\tt add} or {\tt remove} functions. Thus, the patch semantic distribution from both {\tt CC2Vec} and {\tt CCRep} is mixed with {\tt add} and {\tt update}.

\find{{\bf \ding{45} Answer to RQ-1.3: }$\blacktriangleright$ 
Compared with existing patch representation models, \tool is more effective in detecting the intention of patches.
$\blacktriangleleft$ }



\subsection{[RQ-2]: Ablation Study}
[\textbf{Experiment Goal}]: 
We perform an ablation study to investigate the effectiveness of each component in \tool. The major novelty of \tool is the fact that it explicitly includes and processes: \ding{182} {\em SeqIntention} represents intention embedding of the patch at the sequential level, and \ding{183} {\em GraphIntention} represents intention embedding of the patch at the structural level. 

\noindent
[\textbf{Experiment Design}]: We investigate the related contribution of {\em SeqIntention} and {\em GraphIntention} by building two variants of \linebreak \tool where we remove either  {\em GraphIntention}  (\ie denoted as \tool$_{GraphIntention}$), or  {\em SeqIntention} (\ie denoted as \linebreak \tool$_{SeqIntention}$). We also build a native model by removing both {\em GraphIntention} and {\em SeqIntention} components (\ie denoted as \tool$_{both-}$ for comparison. We evaluate the performance of these variants on the task of patch description generation.



\noindent
\textbf{[Experiment Results (RQ-2)]:}
Table \ref{tab:ablation} summarizes the results of our ablation test on the three variants of \tool.
\vspace{1mm}
\begin{table}[h]
    \centering
    \caption{Ablation study results based on the patch description generation task.}
  \resizebox{0.8\linewidth}{!}{
  \begin{tabular}{lrrr}
    \toprule
    Model & ROUGE-L (\%) &BLEU (\%) &  METEOR (\%)\\
    \midrule
    \tool${GraphIntention-} $ & 20.10 & 16.50& 15.40 \\
    \tool${SeqIntention-} $ & 18.44 & 14.70& 16.20 \\
    \tool$_{both-}$ & 15.00 & 13.00 & 12.00 \\
    \midrule
    \tool   & 25.45&23.52&  21.23\\
    \bottomrule
    \end{tabular}
}
    \label{tab:ablation}
\end{table}

While the performance of \tool is not the simple addition of the performance of each variant, we note with \tool$_{both-}$ that the performance is quasi-insignificant, which means that, put together, both design choices are instrumental for the superior performance of \tool.

\noindent
{\em Contribution of Graph Intention Encoding}: 
We observe that the graph intention embedding significantly improves the model ability to generate correct patch descriptions for more patches which is evidenced by the large improvement on the ROUGE-L score (from 20.10 to 25.45), where ROUGE-L is recall oriented.

We postulate that even when token sequences (e.g., identifier names) are different among patches, the similarity of the intention graph helps the model to learn that these patches have the same intent. Nevertheless, precision in description generation (i.e., how many words are correct) is highly dependent on the model's ability to generate the exact correct tokens, which is more guaranteed by the context and sequence intention embedding.



We manually checked different samples to analyze how the variants were performing. Figure~\ref{fig:9} presents a real-world case in our dataset, including the patch, the ground truth, and the patch descriptions generated by \tool, \tool$_{GraphIntention-}$, \tool$_{SeqIntention-}$, as well as three of the strongest baselines for this task (\ie CC2Vec, FIRA, and CCRep). 
In this case, the embeddings of \tool and \tool$_{GraphIntention-}$ are effective in spotting the sub-token {\em trident} in class name {\em TridentTopologyBuilder} thanks to {\tt BPE}. 
In addition, \tool takes advantage of both the sequence intention and graph intention inside the patch.
However, if we only consider the graph intention, \tool$_{SeqIntention-}$ performs the worst against \tool$_{GraphIntention-}$. From the example, we find that {\tt CC2Vec}, which is retrieval-based, cannot generate a proper message because there may not exist similar patches in the training set. {\tt FIRA}, while underperforming against \tool, still performs relatively well because it uses the edition operation detector and sequential contextual information.

\begin{figure}[H]
    \centering
    \begin{tabular}{c}
\begin{lstlisting}[language=diff,linewidth={0.95\linewidth},frame=single]
--- a/src/[...]/topology/TridentTopologyBuilder.java  
+++ b/src/[...]/topology/TridentTopologyBuilder.java
public class TridentTopologyBuilder {
	bd.allGrouping( masterCoordinator( batchGroup ) , MasterBatchCoordinator.
	COMMIT_STREAM_ID);
	for(Map m : c.componentConfs) { 
- 		scd . addConfigurations ( m ) ; 
+ 		bd . addConfigurations ( m ) ; 
	}
}}
\end{lstlisting}
    \end{tabular}

   \vspace{.1em}
   \resizebox{.9\linewidth}{!}
   {\begin{tabular}{ll}
    \toprule 
    Source & Patch description\\\midrule
    \bf Ground truth & \bf set component configurations correctly for trident spouts \\
    \tool & set component configurations correctly for trident spouts\\
    \tool$_{GraphIntention}-$ & configure components for trident \\
    \tool$_{SeqIntention}-$ & set trident components. \\
    CC2Vec & fixed flickering in the preview pane in refactoring preview \\
    FIRA & use the correct component content in onesidediffviewer \\
    CCRep & update for function \\
    \bottomrule
    \end{tabular}}
    \caption{Case analysis of the ablation study}
    \label{fig:9}
\end{figure}


It is noteworthy that \tool is able to generate the token {\em spouts}. This is not due to data leakage since the ground truth commit message was not part of the training set. 
However, our approach builds on a dictionary that considers all tokens in the dataset (just as the entire English dictionary would be considered in text generation). Hence {\em spouts} was predicted from the dictionary as the most probable 
 (using {\em softmax}) token to generate after {\em trident}.

\vspace{-2mm}
\find{{\bf \ding{45} Answer to RQ-2: }$\blacktriangleright$ 
Evaluations of individual components of \tool indicate that both {\em GraphIntention} and {\em SeqIntention} bring a significant portion of the performance. 
$\blacktriangleleft$ }



\vspace{-4mm}
\subsection{[RQ-3]: Robustness Evaluation}
[\textbf{Experiment Design}]: We evaluate the robustness of \tool and of  state-of-the-art patch representation techniques (\ie CC2Vec \cite{hoang2020cc2vec} and CCRep~\cite{liu2023ccrep}) on an independent dataset for the task of patch description generation.
To that end, we pre-train \tool on the dataset used for patch description generation task~\ref{sec:dataset}, but we use, for testing, the dataset of patches that was collected for  patch correctness assessment. The test dataset is consequently independent as the patches come from a different set of projects. We indeed confirm  that the samples across the two datasets are substantially different: Table~\ref{tab:statistics} depicts some basic statistics, which reveal that the average patch sizes and commit message sizes are different. A comparison of patch examples shows that patches in the patch description generation task dataset are small (mostly few-line changes as in Figure~\ref{fig:9}), while the patches in the independent dataset are larger (cf. Figure~\ref{fig:4777}). Conversely, the description messages are an order of magnitude shorter in the independent dataset than in the patch description task dataset.
\begin{table}[h!]
    \centering
    \caption{Dataset statistics}
    \resizebox{0.75\width}{!}{
    \begin{tabular}{lrr}
    \toprule
    & \makecell[r]{patch description\\ generation} & \makecell[r]{patch correctness\\ validation} \\
    \midrule
    Avg. size of patches (\#tokens) & 43.70 &217.98\\
    Avg. size of descriptions (\#tokens)  & 6.97& --\\
    
    \bottomrule
    \end{tabular}}
    \label{tab:statistics}
\end{table}

    

Following the IR-based techniques to retrieve messages for given patches, we use the generated embeddings to match highly similar test set patches with training set patches. 
Then, we select the descriptions associated with the matched training patches  as the "generated" (actually retrieved) descriptions.
 
\begin{figure}[t]
    \centering
    \begin{tabular}{c}
\begin{lstlisting}[language=diff,linewidth={0.95\linewidth},frame=single]
--- a/src/[...]/math/analysis/solvers/BaseSecantSolver.java 
+++ b/src/[...]/math/analysis/solvers/BaseSecantSolver.java 
@@ -183,14 +183,7 @@ public abstract class BaseSecantSolver 
	f0 *= f1 / (f1 + fx); break; 
	case REGULA_FALSI: 
-   if (x == x1) { 
-   	final double delta = FastMath.max(...), 
+// Nothing. 
	break; 
	...
\end{lstlisting}
    \end{tabular}
    
   \resizebox{.85\linewidth}{!}{
   \begin{tabular}{ll}
    \toprule
    Source & Patch description\\\midrule
    \textbf{Ground truth} & \textbf{Remove an obsolete code line} \\
    CC2Vec~\cite{hoang2020cc2vec} & wrap the root cause rather than just using the message \\
    CCRep~\cite{liu2023ccrep} & add notes \\
    \tool & remove obsolete code \\
    \bottomrule
    \end{tabular}}
    \caption{Case analysis: robustness evaluation}
    \label{fig:4777}
\end{figure}

\noindent
\textbf{[Experiment Results (RQ-3)]:}
Table~\ref{tab:rq3} summarizes the experimental results of \tool, CC2Vec, and CCRep baselines on the independent dataset. Compared to CCRep, \tool improves the score from 19.65\% to 21.64\%, 23.67\% to 31.92\%, 12.77\% to 15.37\% for metrics BLEU, ROUGE-L and METEOR, respectively. 
\begin{table}[H]
    \centering
    \caption{Results on independent dataset (\%)}
     \resizebox{0.8\width}{!}{
    \begin{tabular}{lrrr}
    \toprule
     Model & ROUGE-L & BLEU &  METEOR \\ 
     \midrule
    CC2Vec~\cite{hoang2020cc2vec} & 17.34 &9.20&  5.14\\
    CCRep~\cite{liu2023ccrep} & 23.67& 19.65 & 12.77 \\
    \tool & 31.92 &21.64& 15.37\\
     \bottomrule
    \end{tabular}
     }
    \label{tab:rq3}
\end{table}

Figure \ref{fig:4777} illustrates a case where \tool performs better than both {\tt CC2Vec} and {\tt CCRep}. \tool indeed manages to match and return a correct message from the retrieval target dataset. In contrast, CC2Vec fails to retrieve the precise patch description and CCRep obtains inaccurate query back. Such observations further confirm our intuition that {\em SeqIntention} and {\em GraphIntention} can help the model capture the semantics of the patch independently of the training task dataset.


\find{{\bf \ding{45} Answer to RQ-3: }$\blacktriangleright$ 
Compared to CC2Vec and CCRep on the task of patch description generation on an independent dataset, \tool shows its capability to capture the semantics of patches. \tool achieves 10.13\%, 34.85\% and 20.36\% improvement on CCRep  for ROUGE-L, BLEU and METEOR metrics, respectively.
$\blacktriangleleft$ }

\section{Discussion}\label{sec:discussion} 



\subsection{Threats to Validity}
Threats to internal validity refer to errors in the implementation of compared techniques and our approach. To reduce these threats, in each task, we directly reuse the implementation of the baselines from their reproducible packages whenever available. Otherwise, we re-implement the techniques strictly following their papers. Furthermore, we also build our approach based on existing mature tools/libraries, such as javalang~\cite{thunes2013javalang} for parsing ASTs. 

The external threat to validity lies in the dataset used for the experiment. To mitigate this threat, we build a well-established dataset,  which is a rewritten version  based on datasets from prior works~\cite{hoang2020cc2vec,nie2021coregen,tian2022change}.

The construct threat involves the metrics used in evaluations. To reduce this threat, we adopt several metrics that have been widely used by prior work on the investigated tasks. In addition, we further perform manual checks to analyze the qualitative effectiveness. 

\subsection{Limitations}
First, since \tool relies on {\em SeqIntention} and {\em GraphIntention}, our approach would be less effective when patches cannot be parsed into valid AST graph. In this case, \tool would only take contextual information and {\em SeqIntention} as sources to yield the embeddings. However, this limitation lies only when we cannot access source code repositories in which patches have been committed.

Second, for the patch description generation task, we consider two variants: generation-based and retrieval-based. Normally, we collect datasets by following fixed rules, which leads to the training set containing highly-similar patches with the test set. In this case, generate-based \tool could be less effective than an IR-based approach. Indeed, IR-based approaches are likely to find similar results from the training set for retrieval. Nevertheless, as shown in Table~\ref{tab:T1D2}, even in retrieval-based mode, \tool outperforms the baselines. 

Third, when a given patch contains tokens that are absent from  both vocabularies of patches and messages, \tool will fail to generate or recognize these tokens for all tasks.




\section{Related Work}\label{sec:related}
\subsection{Patch Representation}
There are many studies on the representation of code-like texts, including source code representation \cite{feng2020codebert, elnaggar2021codetrans} and patch representation \cite{hoang2020cc2vec}. Previous works focus on representing given patches into latent distributed vectors. Allamanis~\etal~\cite{allamanis2018survey} propose a comprehensive survey on representation learning of code-like texts.

The existing works on representing code-like texts can be categorized as control-flow graph \cite{defreez2018path}, and deep-learning approaches \cite{elnaggar2021codetrans, feng2020codebert, hoang2020cc2vec}. Before learning distributed representations, Henkel~\etal~\cite{henkel2018code} proposes a toolchain to produce abstracted intra-procedural symbolic traces for learning word representations. They conducted their experiments on a downstream task to find and repair bugs in incorrect codes. Wang~\etal~\cite{wang2017dynamic} learns embeddings of code-like text by the usage of execution traces. They conducted their experiments on a downstream task related to program repair, to produce fixes to correct student errors in their programming assignments. 

To leverage deep learning models, Hoang~\etal~\cite{hoang2020cc2vec} proposed CC2Vec, a sequence learning-based approach to represent patches and conduct experiments on three downstream patch tasks: patch description generation, bug fixing patch identification, and just-in-time defect prediction. 
Similarly, CoDiSum~\cite{xu2019commit} is also a token based approach for patch representation that has been used for generating patch descriptions. CCRep~\cite{liu2023ccrep} is an approach to learning code change representations, encoding code changes into feature vectors for a variety of tasks by utilizing pre-trained code models, contextually embedding code, and employing a mechanism called "query back" to extract, encode, and interact with changed code fragments.
Our work improves on these approaches by leveraging the context around the code change and a novel graph intention embedding.
CACHE~\cite{lin2022context} uses AST embeddings for the code change and its surrounding context to learn patch representation for patch correctness prediction.

The closest to our work is FIRA~\cite{dong2022fira} for learning patch descriptions. It uses a special kind of graph that combines the two ASTs before and after the patch with extra special nodes to highlight the relationship (\eg match, add, delete) between the nodes from the two ASTs. Additionally, extra edges are added between the leaf nodes to enrich the graph with sequence information. 
Our work is different is many aspects. 
First, \tool represents the sequence intention and graph intention separately instead of sequence or ASTs, and then learns two different embeddings before combining them.
Second, such representation enables us to leverage powerful SOTA models, \eg Transformer for sequence learning and GCN for graph-based learning.
Third, our GraphIntention representation focuses on learning an embedding of intention of graph changes between AST graphs before and after patching, and not the entire AST which enables the neural model to focus on learning the structural changes. 
Finally, our approach is task-agnostic and can easily be fine-tuned for any patch-related down stream tasks. We have evaluated it on three different tasks while FIRA was only assessed on patch description generation.


\subsection{Applications of Patch Embeddings}
\noindent
\textbf{Patch description generation:} As found by prior studies~\cite{dyer2013boa,dong2022fira}, about 14\% commit messages in 23K java projects are empty. Yet patch description is very significant to developers as they help to quickly understand the intention of the patch without requiring reviewing the entire code. Techniques for patch description generation can be categorized as template-based, information-retrieval-based (IR-based), and generation-based approaches. 
Template-based techniques~\cite{buse2010automatically,cortes2014automatically} analyze the patch and get its correct change type, then generate messages with pre-defined patterns. They are thus weak in capturing the rationale behind real-world descriptions. 
IR-based approaches~\cite{hoang2020cc2vec,liu2018neural,huang2020learning} leverage IR techniques to recall descriptions of the most similar patches from the train set and output them as the "generated" descriptions for the test patches. They generally fail when there is no similar patch between the train set and the test set. 
Generation-based techniques \cite{dong2022fira, xu2019commit, liu2020atom, nie2021coregen} try to learn the semantics of edit operations for patch description generation. Existing such approaches do not account for the bimodal nature of patches (\ie sequence and structure), hence losing the semantics either from the sequential order information or from the semantic logic in the structural abstract syntax trees. With \tool, in order to capture sufficient semantics for patches, we take advantage of both by fusing {\em SeqIntention} and {\em GraphIntention}.

\noindent
\textbf{Patch correctness:}
The state-of-the-art automated program repair techniques mainly rely on the test suite to validate generated patches. Given the weakness of test suites, validated patches are actually only plausible since they can still be incorrect~\cite{qi2015analysis, tian2022predicting, gao2021beyond, gissurarson2022propr, tian2020evaluating, ghanbari2022patch}, due to overfitting.  The research community is therefore investigating efficient methods of automating patch correctness assessment. While some good results can be achieved with dynamic methods~\cite{shariffdeen2021concolic}, static methods are more scalable. Recently, Tian~{\em at al.}~\cite{tian2022best} proposed {\em Panther}, which explored the feasibility of comparing overfitting and correct patches through representation learning techniques (e.g., CC2Vec~\cite{hoang2020cc2vec} and Bert~\cite{devlin2018bert}). 
We show in this work that the representations yielded by \tool can 
vastly improve the results yielded by Panther compared to its current representation learning approaches. 




\section{Conclusion}\label{sec:conclusion}
We present \tool, a novel distributed patch representation learning  approach, which fuses contextual, structural, and sequential information in code changes. In \tool, we model sequential information by the Sequence Intention Encoder to give the model the ability to capture contextual sequence semantics and the sequential intention of patches. In addition, we model structural information by the Graph Intention Encoder to obtain the structural change semantics. Sequence Intention Encoder and Graph Intention Encoder enable \tool to learn high-quality patch representations.

We evaluate \tool on three tasks, and the results demonstrate that it outperforms several baselines, including the state-of-the-art, by substantial margins. An ablation study further highlights the importance of the different design choices. Finally, we compare the robustness of \tool vs the CC2Vec and CCRep state-of-the-art patch representation approach on an independent dataset. The empirical result shows that \tool is more effective.

\textbf{Data Availability:}
We make our code and dataset publicly available at:
\begin{center}
      {\url{https://anonymous.4open.science/r/Patcherizer-1E04}}  
\end{center}

\balance
\bibliographystyle{ACM-Reference-Format}
\bibliography{sample-base}


\begin{thebibliography}{79}


\ifx \showCODEN    \undefined \def \showCODEN     #1{\unskip}     \fi
\ifx \showDOI      \undefined \def \showDOI       #1{#1}\fi
\ifx \showISBNx    \undefined \def \showISBNx     #1{\unskip}     \fi
\ifx \showISBNxiii \undefined \def \showISBNxiii  #1{\unskip}     \fi
\ifx \showISSN     \undefined \def \showISSN      #1{\unskip}     \fi
\ifx \showLCCN     \undefined \def \showLCCN      #1{\unskip}     \fi
\ifx \shownote     \undefined \def \shownote      #1{#1}          \fi
\ifx \showarticletitle \undefined \def \showarticletitle #1{#1}   \fi
\ifx \showURL      \undefined \def \showURL       {\relax}        \fi
\providecommand\bibfield[2]{#2}
\providecommand\bibinfo[2]{#2}
\providecommand\natexlab[1]{#1}
\providecommand\showeprint[2][]{arXiv:#2}

\bibitem[net(2018)]%
        {networkx}
 \bibinfo{year}{2018}\natexlab{}.
\newblock \showarticletitle{NetworkX}.
\newblock In \bibinfo{booktitle}{\emph{Encyclopedia of Social Network Analysis
  and Mining, 2nd Edition}}, \bibfield{editor}{\bibinfo{person}{Reda Alhajj}
  {and} \bibinfo{person}{Jon~G. Rokne}} (Eds.). \bibinfo{publisher}{Springer}.
\newblock
\urldef\tempurl%
\url{https://doi.org/10.1007/978-1-4939-7131-2\_100771}
\showDOI{\tempurl}


\bibitem[Allamanis et~al\mbox{.}(2018)]%
        {allamanis2018survey}
\bibfield{author}{\bibinfo{person}{Miltiadis Allamanis},
  \bibinfo{person}{Earl~T Barr}, \bibinfo{person}{Premkumar Devanbu}, {and}
  \bibinfo{person}{Charles Sutton}.} \bibinfo{year}{2018}\natexlab{}.
\newblock \showarticletitle{A survey of machine learning for big code and
  naturalness}.
\newblock \bibinfo{journal}{\emph{ACM Computing Surveys (CSUR)}}
  \bibinfo{volume}{51}, \bibinfo{number}{4} (\bibinfo{year}{2018}),
  \bibinfo{pages}{1--37}.
\newblock


\bibitem[Alon et~al\mbox{.}(2020)]%
        {alonanycodegen}
\bibfield{author}{\bibinfo{person}{Uri Alon}, \bibinfo{person}{Roy Sadaka},
  \bibinfo{person}{Omer Levy}, {and} \bibinfo{person}{Eran Yahav}.}
  \bibinfo{year}{2020}\natexlab{}.
\newblock \showarticletitle{Structural Language Models of Code}. In
  \bibinfo{booktitle}{\emph{Proceedings of the 37th International Conference on
  Machine Learning, {ICML} 2020, 13-18 July 2020, Virtual Event}}
  \emph{(\bibinfo{series}{Proceedings of Machine Learning Research},
  Vol.~\bibinfo{volume}{119})}. \bibinfo{publisher}{{PMLR}},
  \bibinfo{pages}{245--256}.
\newblock
\urldef\tempurl%
\url{http://proceedings.mlr.press/v119/alon20a.html}
\showURL{%
\tempurl}


\bibitem[Alon et~al\mbox{.}(2019)]%
        {Alon2019}
\bibfield{author}{\bibinfo{person}{Uri Alon}, \bibinfo{person}{Meital
  Zilberstein}, \bibinfo{person}{Omer Levy}, {and} \bibinfo{person}{Eran
  Yahav}.} \bibinfo{year}{2019}\natexlab{}.
\newblock \showarticletitle{code2vec: learning distributed representations of
  code}.
\newblock \bibinfo{journal}{\emph{{PACMPL}}} \bibinfo{volume}{3},
  \bibinfo{number}{{POPL}} (\bibinfo{year}{2019}),
  \bibinfo{pages}{40:1--40:29}.
\newblock
\urldef\tempurl%
\url{https://doi.org/10.1145/3290353}
\showDOI{\tempurl}


\bibitem[Banerjee and Lavie(2005)]%
        {banerjee2005meteor}
\bibfield{author}{\bibinfo{person}{Satanjeev Banerjee} {and}
  \bibinfo{person}{Alon Lavie}.} \bibinfo{year}{2005}\natexlab{}.
\newblock \showarticletitle{METEOR: An automatic metric for MT evaluation with
  improved correlation with human judgments}. In
  \bibinfo{booktitle}{\emph{Proceedings of the acl workshop on intrinsic and
  extrinsic evaluation measures for machine translation and/or summarization}}.
  \bibinfo{pages}{65--72}.
\newblock


\bibitem[Barr et~al\mbox{.}(2014)]%
        {barr2014plastic}
\bibfield{author}{\bibinfo{person}{Earl~T Barr}, \bibinfo{person}{Yuriy Brun},
  \bibinfo{person}{Premkumar Devanbu}, \bibinfo{person}{Mark Harman}, {and}
  \bibinfo{person}{Federica Sarro}.} \bibinfo{year}{2014}\natexlab{}.
\newblock \showarticletitle{The plastic surgery hypothesis}. In
  \bibinfo{booktitle}{\emph{Proceedings of the 22nd ACM SIGSOFT International
  Symposium on Foundations of Software Engineering}}.
  \bibinfo{pages}{306--317}.
\newblock


\bibitem[Brody et~al\mbox{.}(2020)]%
        {brody2020structural}
\bibfield{author}{\bibinfo{person}{Shaked Brody}, \bibinfo{person}{Uri Alon},
  {and} \bibinfo{person}{Eran Yahav}.} \bibinfo{year}{2020}\natexlab{}.
\newblock \showarticletitle{A structural model for contextual code changes}.
\newblock \bibinfo{journal}{\emph{Proceedings of the ACM on Programming
  Languages}} \bibinfo{volume}{4}, \bibinfo{number}{OOPSLA}
  (\bibinfo{year}{2020}), \bibinfo{pages}{1--28}.
\newblock


\bibitem[Buse and Weimer(2010)]%
        {buse2010automatically}
\bibfield{author}{\bibinfo{person}{Raymond~PL Buse} {and}
  \bibinfo{person}{Westley~R Weimer}.} \bibinfo{year}{2010}\natexlab{}.
\newblock \showarticletitle{Automatically documenting program changes}. In
  \bibinfo{booktitle}{\emph{Proceedings of the IEEE/ACM international
  conference on Automated software engineering}}. \bibinfo{pages}{33--42}.
\newblock


\bibitem[Cabrera~Lozoya et~al\mbox{.}(2021)]%
        {cabrera2021commit2vec}
\bibfield{author}{\bibinfo{person}{Roc{\'\i}o Cabrera~Lozoya},
  \bibinfo{person}{Arnaud Baumann}, \bibinfo{person}{Antonino Sabetta}, {and}
  \bibinfo{person}{Michele Bezzi}.} \bibinfo{year}{2021}\natexlab{}.
\newblock \showarticletitle{Commit2vec: Learning distributed representations of
  code changes}.
\newblock \bibinfo{journal}{\emph{SN Computer Science}} \bibinfo{volume}{2},
  \bibinfo{number}{3} (\bibinfo{year}{2021}), \bibinfo{pages}{1--16}.
\newblock


\bibitem[Ciniselli et~al\mbox{.}(2021)]%
        {ciniselli2021empirical}
\bibfield{author}{\bibinfo{person}{Matteo Ciniselli}, \bibinfo{person}{Nathan
  Cooper}, \bibinfo{person}{Luca Pascarella}, \bibinfo{person}{Denys
  Poshyvanyk}, \bibinfo{person}{Massimiliano Di~Penta}, {and}
  \bibinfo{person}{Gabriele Bavota}.} \bibinfo{year}{2021}\natexlab{}.
\newblock \showarticletitle{An empirical study on the usage of BERT models for
  code completion}. In \bibinfo{booktitle}{\emph{2021 IEEE/ACM 18th
  International Conference on Mining Software Repositories (MSR)}}. IEEE,
  \bibinfo{pages}{108--119}.
\newblock


\bibitem[Cordella et~al\mbox{.}(1999)]%
        {cordella1999performance}
\bibfield{author}{\bibinfo{person}{Luigi~P Cordella}, \bibinfo{person}{Pasquale
  Foggia}, \bibinfo{person}{Carlo Sansone}, {and} \bibinfo{person}{Mario
  Vento}.} \bibinfo{year}{1999}\natexlab{}.
\newblock \showarticletitle{Performance evaluation of the VF graph matching
  algorithm}. In \bibinfo{booktitle}{\emph{Proceedings 10th International
  Conference on Image Analysis and Processing}}. IEEE,
  \bibinfo{pages}{1172--1177}.
\newblock


\bibitem[Cort{\'e}s-Coy et~al\mbox{.}(2014)]%
        {cortes2014automatically}
\bibfield{author}{\bibinfo{person}{Luis~Fernando Cort{\'e}s-Coy},
  \bibinfo{person}{Mario Linares-V{\'a}squez}, \bibinfo{person}{Jairo Aponte},
  {and} \bibinfo{person}{Denys Poshyvanyk}.} \bibinfo{year}{2014}\natexlab{}.
\newblock \showarticletitle{On automatically generating commit messages via
  summarization of source code changes}. In \bibinfo{booktitle}{\emph{2014 IEEE
  14th International Working Conference on Source Code Analysis and
  Manipulation}}. IEEE, \bibinfo{pages}{275--284}.
\newblock


\bibitem[DeFreez et~al\mbox{.}(2018)]%
        {defreez2018path}
\bibfield{author}{\bibinfo{person}{Daniel DeFreez}, \bibinfo{person}{Aditya~V
  Thakur}, {and} \bibinfo{person}{Cindy Rubio-Gonz{\'a}lez}.}
  \bibinfo{year}{2018}\natexlab{}.
\newblock \showarticletitle{Path-based function embedding and its application
  to error-handling specification mining}. In
  \bibinfo{booktitle}{\emph{Proceedings of the 2018 26th ACM Joint Meeting on
  European Software Engineering Conference and Symposium on the Foundations of
  Software Engineering}}. \bibinfo{pages}{423--433}.
\newblock


\bibitem[Devlin et~al\mbox{.}(2018)]%
        {devlin2018bert}
\bibfield{author}{\bibinfo{person}{Jacob Devlin}, \bibinfo{person}{Ming-Wei
  Chang}, \bibinfo{person}{Kenton Lee}, {and} \bibinfo{person}{Kristina
  Toutanova}.} \bibinfo{year}{2018}\natexlab{}.
\newblock \showarticletitle{Bert: Pre-training of deep bidirectional
  transformers for language understanding}.
\newblock \bibinfo{journal}{\emph{arXiv preprint arXiv:1810.04805}}
  (\bibinfo{year}{2018}).
\newblock


\bibitem[Dong et~al\mbox{.}(2022)]%
        {dong2022fira}
\bibfield{author}{\bibinfo{person}{Jinhao Dong}, \bibinfo{person}{Yiling Lou},
  \bibinfo{person}{Qihao Zhu}, \bibinfo{person}{Zeyu Sun},
  \bibinfo{person}{Zhilin Li}, \bibinfo{person}{Wenjie Zhang}, {and}
  \bibinfo{person}{Dan Hao}.} \bibinfo{year}{2022}\natexlab{}.
\newblock \showarticletitle{FIRA: Fine-Grained Graph-Based Code Change
  Representation for Automated Commit Message Generation}.
\newblock  (\bibinfo{year}{2022}).
\newblock


\bibitem[Dyer et~al\mbox{.}(2013)]%
        {dyer2013boa}
\bibfield{author}{\bibinfo{person}{Robert Dyer}, \bibinfo{person}{Hoan~Anh
  Nguyen}, \bibinfo{person}{Hridesh Rajan}, {and} \bibinfo{person}{Tien~N
  Nguyen}.} \bibinfo{year}{2013}\natexlab{}.
\newblock \showarticletitle{Boa: A language and infrastructure for analyzing
  ultra-large-scale software repositories}. In \bibinfo{booktitle}{\emph{2013
  35th International Conference on Software Engineering (ICSE)}}. IEEE,
  \bibinfo{pages}{422--431}.
\newblock


\bibitem[Elnaggar et~al\mbox{.}(2021)]%
        {elnaggar2021codetrans}
\bibfield{author}{\bibinfo{person}{Ahmed Elnaggar}, \bibinfo{person}{Wei Ding},
  \bibinfo{person}{Llion Jones}, \bibinfo{person}{Tom Gibbs},
  \bibinfo{person}{Tamas Feher}, \bibinfo{person}{Christoph Angerer},
  \bibinfo{person}{Silvia Severini}, \bibinfo{person}{Florian Matthes}, {and}
  \bibinfo{person}{Burkhard Rost}.} \bibinfo{year}{2021}\natexlab{}.
\newblock \showarticletitle{CodeTrans: Towards Cracking the Language of
  Silicon's Code Through Self-Supervised Deep Learning and High Performance
  Computing}.
\newblock \bibinfo{journal}{\emph{arXiv preprint arXiv:2104.02443}}
  (\bibinfo{year}{2021}).
\newblock


\bibitem[Feng et~al\mbox{.}(2020)]%
        {feng2020codebert}
\bibfield{author}{\bibinfo{person}{Zhangyin Feng}, \bibinfo{person}{Daya Guo},
  \bibinfo{person}{Duyu Tang}, \bibinfo{person}{Nan Duan},
  \bibinfo{person}{Xiaocheng Feng}, \bibinfo{person}{Ming Gong},
  \bibinfo{person}{Linjun Shou}, \bibinfo{person}{Bing Qin},
  \bibinfo{person}{Ting Liu}, \bibinfo{person}{Daxin Jiang}, {et~al\mbox{.}}}
  \bibinfo{year}{2020}\natexlab{}.
\newblock \showarticletitle{Codebert: A pre-trained model for programming and
  natural languages}.
\newblock \bibinfo{journal}{\emph{arXiv preprint arXiv:2002.08155}}
  (\bibinfo{year}{2020}).
\newblock


\bibitem[Gao et~al\mbox{.}(2021)]%
        {gao2021beyond}
\bibfield{author}{\bibinfo{person}{Xiang Gao}, \bibinfo{person}{Bo Wang},
  \bibinfo{person}{Gregory~J Duck}, \bibinfo{person}{Ruyi Ji},
  \bibinfo{person}{Yingfei Xiong}, {and} \bibinfo{person}{Abhik Roychoudhury}.}
  \bibinfo{year}{2021}\natexlab{}.
\newblock \showarticletitle{Beyond tests: Program vulnerability repair via
  crash constraint extraction}.
\newblock \bibinfo{journal}{\emph{ACM Transactions on Software Engineering and
  Methodology (TOSEM)}} \bibinfo{volume}{30}, \bibinfo{number}{2}
  (\bibinfo{year}{2021}), \bibinfo{pages}{1--27}.
\newblock


\bibitem[Ghanbari and Marcus(2022)]%
        {ghanbari2022patch}
\bibfield{author}{\bibinfo{person}{Ali Ghanbari} {and} \bibinfo{person}{Andrian
  Marcus}.} \bibinfo{year}{2022}\natexlab{}.
\newblock \bibinfo{title}{Patch Correctness Assessment in Automated Program
  Repair Based on the Impact of Patches on Production and Test Code}.
\newblock
\newblock


\bibitem[Gissurarson et~al\mbox{.}(2022)]%
        {gissurarson2022propr}
\bibfield{author}{\bibinfo{person}{Matth{\'\i}as~P{\'a}ll Gissurarson},
  \bibinfo{person}{Leonhard Applis}, \bibinfo{person}{Annibale Panichella},
  \bibinfo{person}{Arie van Deursen}, {and} \bibinfo{person}{David Sands}.}
  \bibinfo{year}{2022}\natexlab{}.
\newblock \showarticletitle{PropR: property-based automatic program repair}. In
  \bibinfo{booktitle}{\emph{Proceedings of the 44th International Conference on
  Software Engineering}}. \bibinfo{pages}{1768--1780}.
\newblock


\bibitem[Glorot and Bengio(2010)]%
        {glorot2010understanding}
\bibfield{author}{\bibinfo{person}{Xavier Glorot} {and} \bibinfo{person}{Yoshua
  Bengio}.} \bibinfo{year}{2010}\natexlab{}.
\newblock \showarticletitle{Understanding the difficulty of training deep
  feedforward neural networks}. In \bibinfo{booktitle}{\emph{Proceedings of the
  thirteenth international conference on artificial intelligence and
  statistics}}. JMLR Workshop and Conference Proceedings,
  \bibinfo{pages}{249--256}.
\newblock


\bibitem[Glorot et~al\mbox{.}(2011)]%
        {glorot2011deep}
\bibfield{author}{\bibinfo{person}{Xavier Glorot}, \bibinfo{person}{Antoine
  Bordes}, {and} \bibinfo{person}{Yoshua Bengio}.}
  \bibinfo{year}{2011}\natexlab{}.
\newblock \showarticletitle{Deep sparse rectifier neural networks}. In
  \bibinfo{booktitle}{\emph{Proceedings of the fourteenth international
  conference on artificial intelligence and statistics}}. JMLR Workshop and
  Conference Proceedings, \bibinfo{pages}{315--323}.
\newblock


\bibitem[Guo et~al\mbox{.}(2021)]%
        {DBLP:conf/iclr/GuoRLFT0ZDSFTDC21}
\bibfield{author}{\bibinfo{person}{Daya Guo}, \bibinfo{person}{Shuo Ren},
  \bibinfo{person}{Shuai Lu}, \bibinfo{person}{Zhangyin Feng},
  \bibinfo{person}{Duyu Tang}, \bibinfo{person}{Shujie Liu},
  \bibinfo{person}{Long Zhou}, \bibinfo{person}{Nan Duan},
  \bibinfo{person}{Alexey Svyatkovskiy}, \bibinfo{person}{Shengyu Fu},
  \bibinfo{person}{Michele Tufano}, \bibinfo{person}{Shao~Kun Deng},
  \bibinfo{person}{Colin~B. Clement}, \bibinfo{person}{Dawn Drain},
  \bibinfo{person}{Neel Sundaresan}, \bibinfo{person}{Jian Yin},
  \bibinfo{person}{Daxin Jiang}, {and} \bibinfo{person}{Ming Zhou}.}
  \bibinfo{year}{2021}\natexlab{}.
\newblock \showarticletitle{GraphCodeBERT: Pre-training Code Representations
  with Data Flow}. In \bibinfo{booktitle}{\emph{9th International Conference on
  Learning Representations, {ICLR} 2021, Virtual Event, Austria, May 3-7,
  2021}}. \bibinfo{publisher}{OpenReview.net}.
\newblock
\urldef\tempurl%
\url{https://openreview.net/forum?id=jLoC4ez43PZ}
\showURL{%
\tempurl}


\bibitem[He et~al\mbox{.}(2016)]%
        {he2016deep}
\bibfield{author}{\bibinfo{person}{Kaiming He}, \bibinfo{person}{Xiangyu
  Zhang}, \bibinfo{person}{Shaoqing Ren}, {and} \bibinfo{person}{Jian Sun}.}
  \bibinfo{year}{2016}\natexlab{}.
\newblock \showarticletitle{Deep residual learning for image recognition}. In
  \bibinfo{booktitle}{\emph{Proceedings of the IEEE conference on computer
  vision and pattern recognition}}. \bibinfo{pages}{770--778}.
\newblock


\bibitem[Henkel et~al\mbox{.}(2018)]%
        {henkel2018code}
\bibfield{author}{\bibinfo{person}{Jordan Henkel}, \bibinfo{person}{Shuvendu~K
  Lahiri}, \bibinfo{person}{Ben Liblit}, {and} \bibinfo{person}{Thomas Reps}.}
  \bibinfo{year}{2018}\natexlab{}.
\newblock \showarticletitle{Code vectors: Understanding programs through
  embedded abstracted symbolic traces}. In
  \bibinfo{booktitle}{\emph{Proceedings of the 2018 26th ACM Joint Meeting on
  European Software Engineering Conference and Symposium on the Foundations of
  Software Engineering}}. \bibinfo{pages}{163--174}.
\newblock


\bibitem[Hoang et~al\mbox{.}(2019)]%
        {hoang2019deepjit}
\bibfield{author}{\bibinfo{person}{Thong Hoang}, \bibinfo{person}{Hoa~Khanh
  Dam}, \bibinfo{person}{Yasutaka Kamei}, \bibinfo{person}{David Lo}, {and}
  \bibinfo{person}{Naoyasu Ubayashi}.} \bibinfo{year}{2019}\natexlab{}.
\newblock \showarticletitle{DeepJIT: an end-to-end deep learning framework for
  just-in-time defect prediction}. In \bibinfo{booktitle}{\emph{2019 IEEE/ACM
  16th International Conference on Mining Software Repositories (MSR)}}. IEEE,
  \bibinfo{pages}{34--45}.
\newblock


\bibitem[Hoang et~al\mbox{.}(2020)]%
        {hoang2020cc2vec}
\bibfield{author}{\bibinfo{person}{Thong Hoang}, \bibinfo{person}{Hong~Jin
  Kang}, \bibinfo{person}{David Lo}, {and} \bibinfo{person}{Julia Lawall}.}
  \bibinfo{year}{2020}\natexlab{}.
\newblock \showarticletitle{Cc2vec: Distributed representations of code
  changes}. In \bibinfo{booktitle}{\emph{Proceedings of the ACM/IEEE 42nd
  International Conference on Software Engineering}}.
  \bibinfo{pages}{518--529}.
\newblock


\bibitem[Huang et~al\mbox{.}(2020)]%
        {huang2020learning}
\bibfield{author}{\bibinfo{person}{Yuan Huang}, \bibinfo{person}{Nan Jia},
  \bibinfo{person}{Hao-Jie Zhou}, \bibinfo{person}{Xiang-Ping Chen},
  \bibinfo{person}{Zi-Bin Zheng}, {and} \bibinfo{person}{Ming-Dong Tang}.}
  \bibinfo{year}{2020}\natexlab{}.
\newblock \showarticletitle{Learning human-written commit messages to document
  code changes}.
\newblock \bibinfo{journal}{\emph{Journal of Computer Science and Technology}}
  \bibinfo{volume}{35}, \bibinfo{number}{6} (\bibinfo{year}{2020}),
  \bibinfo{pages}{1258--1277}.
\newblock


\bibitem[Jiang et~al\mbox{.}(2021)]%
        {jiang2021cure}
\bibfield{author}{\bibinfo{person}{Nan Jiang}, \bibinfo{person}{Thibaud
  Lutellier}, {and} \bibinfo{person}{Lin Tan}.}
  \bibinfo{year}{2021}\natexlab{}.
\newblock \showarticletitle{CURE: Code-aware neural machine translation for
  automatic program repair}. In \bibinfo{booktitle}{\emph{2021 IEEE/ACM 43rd
  International Conference on Software Engineering (ICSE)}}. IEEE,
  \bibinfo{pages}{1161--1173}.
\newblock


\bibitem[Jiang et~al\mbox{.}(2017)]%
        {jiang2017automatically}
\bibfield{author}{\bibinfo{person}{Siyuan Jiang}, \bibinfo{person}{Ameer
  Armaly}, {and} \bibinfo{person}{Collin McMillan}.}
  \bibinfo{year}{2017}\natexlab{}.
\newblock \showarticletitle{Automatically generating commit messages from diffs
  using neural machine translation}. In \bibinfo{booktitle}{\emph{Proceedings
  of the 32nd IEEE/ACM International Conference on Automated Software
  Engineering}}. IEEE, \bibinfo{pages}{135--146}.
\newblock


\bibitem[Kamei et~al\mbox{.}(2016)]%
        {kamei2016studying}
\bibfield{author}{\bibinfo{person}{Yasutaka Kamei}, \bibinfo{person}{Takafumi
  Fukushima}, \bibinfo{person}{Shane McIntosh}, \bibinfo{person}{Kazuhiro
  Yamashita}, \bibinfo{person}{Naoyasu Ubayashi}, {and}
  \bibinfo{person}{Ahmed~E Hassan}.} \bibinfo{year}{2016}\natexlab{}.
\newblock \showarticletitle{Studying just-in-time defect prediction using
  cross-project models}.
\newblock \bibinfo{journal}{\emph{Empirical Software Engineering}}
  \bibinfo{volume}{21}, \bibinfo{number}{5} (\bibinfo{year}{2016}),
  \bibinfo{pages}{2072--2106}.
\newblock


\bibitem[Kamei et~al\mbox{.}(2012)]%
        {kamei2012large}
\bibfield{author}{\bibinfo{person}{Yasutaka Kamei}, \bibinfo{person}{Emad
  Shihab}, \bibinfo{person}{Bram Adams}, \bibinfo{person}{Ahmed~E Hassan},
  \bibinfo{person}{Audris Mockus}, \bibinfo{person}{Anand Sinha}, {and}
  \bibinfo{person}{Naoyasu Ubayashi}.} \bibinfo{year}{2012}\natexlab{}.
\newblock \showarticletitle{A large-scale empirical study of just-in-time
  quality assurance}.
\newblock \bibinfo{journal}{\emph{IEEE Transactions on Software Engineering}}
  \bibinfo{volume}{39}, \bibinfo{number}{6} (\bibinfo{year}{2012}),
  \bibinfo{pages}{757--773}.
\newblock


\bibitem[Kingma and Ba(2014)]%
        {kingma2014adam}
\bibfield{author}{\bibinfo{person}{Diederik~P Kingma} {and}
  \bibinfo{person}{Jimmy Ba}.} \bibinfo{year}{2014}\natexlab{}.
\newblock \showarticletitle{Adam: A method for stochastic optimization}.
\newblock \bibinfo{journal}{\emph{arXiv preprint arXiv:1412.6980}}
  (\bibinfo{year}{2014}).
\newblock


\bibitem[Kipf and Welling(2016)]%
        {kipf2016semi}
\bibfield{author}{\bibinfo{person}{Thomas~N Kipf} {and} \bibinfo{person}{Max
  Welling}.} \bibinfo{year}{2016}\natexlab{}.
\newblock \showarticletitle{Semi-supervised classification with graph
  convolutional networks}.
\newblock \bibinfo{journal}{\emph{arXiv preprint arXiv:1609.02907}}
  (\bibinfo{year}{2016}).
\newblock


\bibitem[Li et~al\mbox{.}(2021)]%
        {li2021attention}
\bibfield{author}{\bibinfo{person}{Min Li}, \bibinfo{person}{Zhenjiang Miao},
  \bibinfo{person}{Xiao-Ping Zhang}, {and} \bibinfo{person}{Wanru Xu}.}
  \bibinfo{year}{2021}\natexlab{}.
\newblock \showarticletitle{An Attention-Seq2Seq Model Based on CRNN Encoding
  for Automatic Labanotation Generation from Motion Capture Data}. In
  \bibinfo{booktitle}{\emph{ICASSP 2021-2021 IEEE International Conference on
  Acoustics, Speech and Signal Processing (ICASSP)}}. IEEE,
  \bibinfo{pages}{4185--4189}.
\newblock


\bibitem[Lin et~al\mbox{.}(2022)]%
        {lin2022context}
\bibfield{author}{\bibinfo{person}{Bo Lin}, \bibinfo{person}{Shangwen Wang},
  \bibinfo{person}{Ming Wen}, {and} \bibinfo{person}{Xiaoguang Mao}.}
  \bibinfo{year}{2022}\natexlab{}.
\newblock \showarticletitle{Context-aware code change embedding for better
  patch correctness assessment}.
\newblock \bibinfo{journal}{\emph{ACM Transactions on Software Engineering and
  Methodology (TOSEM)}} \bibinfo{volume}{31}, \bibinfo{number}{3}
  (\bibinfo{year}{2022}), \bibinfo{pages}{1--29}.
\newblock


\bibitem[Linares-V{\'a}squez et~al\mbox{.}(2015)]%
        {linares2015changescribe}
\bibfield{author}{\bibinfo{person}{Mario Linares-V{\'a}squez},
  \bibinfo{person}{Luis~Fernando Cort{\'e}s-Coy}, \bibinfo{person}{Jairo
  Aponte}, {and} \bibinfo{person}{Denys Poshyvanyk}.}
  \bibinfo{year}{2015}\natexlab{}.
\newblock \showarticletitle{Changescribe: A tool for automatically generating
  commit messages}. In \bibinfo{booktitle}{\emph{2015 IEEE/ACM 37th IEEE
  International Conference on Software Engineering}}, Vol.~\bibinfo{volume}{2}.
  IEEE, \bibinfo{pages}{709--712}.
\newblock


\bibitem[Liu et~al\mbox{.}(2020b)]%
        {liu2020self}
\bibfield{author}{\bibinfo{person}{Fang Liu}, \bibinfo{person}{Ge Li},
  \bibinfo{person}{Bolin Wei}, \bibinfo{person}{Xin Xia},
  \bibinfo{person}{Zhiyi Fu}, {and} \bibinfo{person}{Zhi Jin}.}
  \bibinfo{year}{2020}\natexlab{b}.
\newblock \showarticletitle{A self-attentional neural architecture for code
  completion with multi-task learning}. In
  \bibinfo{booktitle}{\emph{Proceedings of the 28th International Conference on
  Program Comprehension}}. \bibinfo{pages}{37--47}.
\newblock


\bibitem[Liu et~al\mbox{.}(2020c)]%
        {liu2020multi}
\bibfield{author}{\bibinfo{person}{Fang Liu}, \bibinfo{person}{Ge Li},
  \bibinfo{person}{Yunfei Zhao}, {and} \bibinfo{person}{Zhi Jin}.}
  \bibinfo{year}{2020}\natexlab{c}.
\newblock \showarticletitle{Multi-task learning based pre-trained language
  model for code completion}. In \bibinfo{booktitle}{\emph{Proceedings of the
  35th IEEE/ACM International Conference on Automated Software Engineering}}.
  \bibinfo{pages}{473--485}.
\newblock


\bibitem[Liu et~al\mbox{.}(2019)]%
        {liu2019generating}
\bibfield{author}{\bibinfo{person}{Qin Liu}, \bibinfo{person}{Zihe Liu},
  \bibinfo{person}{Hongming Zhu}, \bibinfo{person}{Hongfei Fan},
  \bibinfo{person}{Bowen Du}, {and} \bibinfo{person}{Yu Qian}.}
  \bibinfo{year}{2019}\natexlab{}.
\newblock \showarticletitle{Generating commit messages from diffs using
  pointer-generator network}. In \bibinfo{booktitle}{\emph{2019 IEEE/ACM 16th
  International Conference on Mining Software Repositories (MSR)}}. IEEE,
  \bibinfo{pages}{299--309}.
\newblock


\bibitem[Liu et~al\mbox{.}(2020a)]%
        {liu2020atom}
\bibfield{author}{\bibinfo{person}{Shangqing Liu}, \bibinfo{person}{Cuiyun
  Gao}, \bibinfo{person}{Sen Chen}, \bibinfo{person}{Nie~Lun Yiu}, {and}
  \bibinfo{person}{Yang Liu}.} \bibinfo{year}{2020}\natexlab{a}.
\newblock \showarticletitle{ATOM: Commit message generation based on abstract
  syntax tree and hybrid ranking}.
\newblock \bibinfo{journal}{\emph{IEEE Transactions on Software Engineering}}
  (\bibinfo{year}{2020}).
\newblock


\bibitem[Liu et~al\mbox{.}(2023)]%
        {liu2023ccrep}
\bibfield{author}{\bibinfo{person}{Zhongxin Liu}, \bibinfo{person}{Zhijie
  Tang}, \bibinfo{person}{Xin Xia}, {and} \bibinfo{person}{Xiaohu Yang}.}
  \bibinfo{year}{2023}\natexlab{}.
\newblock \showarticletitle{CCRep: Learning Code Change Representations via
  Pre-Trained Code Model and Query Back}. In \bibinfo{booktitle}{\emph{45th
  {IEEE/ACM} International Conference on Software Engineering, {ICSE} 2023,
  Melbourne, Australia, May 14-20, 2023}}. \bibinfo{publisher}{{IEEE}},
  \bibinfo{pages}{17--29}.
\newblock
\urldef\tempurl%
\url{https://doi.org/10.1109/ICSE48619.2023.00014}
\showDOI{\tempurl}


\bibitem[Liu et~al\mbox{.}(2018)]%
        {liu2018neural}
\bibfield{author}{\bibinfo{person}{Zhongxin Liu}, \bibinfo{person}{Xin Xia},
  \bibinfo{person}{Ahmed~E Hassan}, \bibinfo{person}{David Lo},
  \bibinfo{person}{Zhenchang Xing}, {and} \bibinfo{person}{Xinyu Wang}.}
  \bibinfo{year}{2018}\natexlab{}.
\newblock \showarticletitle{Neural-machine-translation-based commit message
  generation: how far are we?}. In \bibinfo{booktitle}{\emph{Proceedings of the
  33rd ACM/IEEE International Conference on Automated Software Engineering}}.
  \bibinfo{pages}{373--384}.
\newblock


\bibitem[Luo et~al\mbox{.}(2018)]%
        {luo2018cosine}
\bibfield{author}{\bibinfo{person}{Chunjie Luo}, \bibinfo{person}{Jianfeng
  Zhan}, \bibinfo{person}{Xiaohe Xue}, \bibinfo{person}{Lei Wang},
  \bibinfo{person}{Rui Ren}, {and} \bibinfo{person}{Qiang Yang}.}
  \bibinfo{year}{2018}\natexlab{}.
\newblock \showarticletitle{Cosine normalization: Using cosine similarity
  instead of dot product in neural networks}. In
  \bibinfo{booktitle}{\emph{International Conference on Artificial Neural
  Networks}}. Springer, \bibinfo{pages}{382--391}.
\newblock


\bibitem[Nie et~al\mbox{.}(2021)]%
        {nie2021coregen}
\bibfield{author}{\bibinfo{person}{Lun~Yiu Nie}, \bibinfo{person}{Cuiyun Gao},
  \bibinfo{person}{Zhicong Zhong}, \bibinfo{person}{Wai Lam},
  \bibinfo{person}{Yang Liu}, {and} \bibinfo{person}{Zenglin Xu}.}
  \bibinfo{year}{2021}\natexlab{}.
\newblock \showarticletitle{CoreGen: Contextualized Code Representation
  Learning for Commit Message Generation}.
\newblock \bibinfo{journal}{\emph{Neurocomputing}}  \bibinfo{volume}{459}
  (\bibinfo{year}{2021}), \bibinfo{pages}{97--107}.
\newblock


\bibitem[Niemeyer and Geiger(2021)]%
        {niemeyer2021giraffe}
\bibfield{author}{\bibinfo{person}{Michael Niemeyer} {and}
  \bibinfo{person}{Andreas Geiger}.} \bibinfo{year}{2021}\natexlab{}.
\newblock \showarticletitle{Giraffe: Representing scenes as compositional
  generative neural feature fields}. In \bibinfo{booktitle}{\emph{Proceedings
  of the IEEE/CVF Conference on Computer Vision and Pattern Recognition}}.
  \bibinfo{pages}{11453--11464}.
\newblock


\bibitem[Papineni et~al\mbox{.}(2002)]%
        {papineni2002bleu}
\bibfield{author}{\bibinfo{person}{Kishore Papineni}, \bibinfo{person}{Salim
  Roukos}, \bibinfo{person}{Todd Ward}, {and} \bibinfo{person}{Wei-Jing Zhu}.}
  \bibinfo{year}{2002}\natexlab{}.
\newblock \showarticletitle{Bleu: a method for automatic evaluation of machine
  translation}. In \bibinfo{booktitle}{\emph{Proceedings of the 40th annual
  meeting of the Association for Computational Linguistics}}.
  \bibinfo{pages}{311--318}.
\newblock


\bibitem[Pian et~al\mbox{.}(2022)]%
        {pian2022metatptrans}
\bibfield{author}{\bibinfo{person}{Weiguo Pian}, \bibinfo{person}{Hanyu Peng},
  \bibinfo{person}{Xunzhu Tang}, \bibinfo{person}{Tiezhu Sun},
  \bibinfo{person}{Haoye Tian}, \bibinfo{person}{Andrew Habib},
  \bibinfo{person}{Jacques Klein}, {and} \bibinfo{person}{Tegawend{\'e}~F
  Bissyand{\'e}}.} \bibinfo{year}{2022}\natexlab{}.
\newblock \showarticletitle{MetaTPTrans: A Meta Learning Approach for
  Multilingual Code Representation Learning}.
\newblock \bibinfo{journal}{\emph{arXiv preprint arXiv:2206.06460}}
  (\bibinfo{year}{2022}).
\newblock


\bibitem[Pian et~al\mbox{.}(2023)]%
        {pian2023metatptrans}
\bibfield{author}{\bibinfo{person}{Weiguo Pian}, \bibinfo{person}{Hanyu Peng},
  \bibinfo{person}{Xunzhu Tang}, \bibinfo{person}{Tiezhu Sun},
  \bibinfo{person}{Haoye Tian}, \bibinfo{person}{Andrew Habib},
  \bibinfo{person}{Jacques Klein}, {and} \bibinfo{person}{Tegawend{\'e}~F
  Bissyand{\'e}}.} \bibinfo{year}{2023}\natexlab{}.
\newblock \showarticletitle{MetaTPTrans: A meta learning approach for
  multilingual code representation learning}. In
  \bibinfo{booktitle}{\emph{Proceedings of the AAAI Conference on Artificial
  Intelligence}}, Vol.~\bibinfo{volume}{37}. \bibinfo{pages}{5239--5247}.
\newblock


\bibitem[Qi et~al\mbox{.}(2015)]%
        {qi2015analysis}
\bibfield{author}{\bibinfo{person}{Zichao Qi}, \bibinfo{person}{Fan Long},
  \bibinfo{person}{Sara Achour}, {and} \bibinfo{person}{Martin Rinard}.}
  \bibinfo{year}{2015}\natexlab{}.
\newblock \showarticletitle{An analysis of patch plausibility and correctness
  for generate-and-validate patch generation systems}. In
  \bibinfo{booktitle}{\emph{Proceedings of the 2015 International Symposium on
  Software Testing and Analysis}}. \bibinfo{pages}{24--36}.
\newblock


\bibitem[Qin et~al\mbox{.}(2021)]%
        {qin2021co}
\bibfield{author}{\bibinfo{person}{Libo Qin}, \bibinfo{person}{Tailu Liu},
  \bibinfo{person}{Wanxiang Che}, \bibinfo{person}{Bingbing Kang},
  \bibinfo{person}{Sendong Zhao}, {and} \bibinfo{person}{Ting Liu}.}
  \bibinfo{year}{2021}\natexlab{}.
\newblock \showarticletitle{A co-interactive transformer for joint slot filling
  and intent detection}. In \bibinfo{booktitle}{\emph{ICASSP 2021-2021 IEEE
  International Conference on Acoustics, Speech and Signal Processing
  (ICASSP)}}. IEEE, \bibinfo{pages}{8193--8197}.
\newblock


\bibitem[ROUGE(2004)]%
        {rouge2004package}
\bibfield{author}{\bibinfo{person}{Lin~CY ROUGE}.}
  \bibinfo{year}{2004}\natexlab{}.
\newblock \showarticletitle{A package for automatic evaluation of summaries}.
  In \bibinfo{booktitle}{\emph{Proceedings of Workshop on Text Summarization of
  ACL, Spain}}.
\newblock


\bibitem[Rubinstein(1999)]%
        {rubinstein1999cross}
\bibfield{author}{\bibinfo{person}{Reuven Rubinstein}.}
  \bibinfo{year}{1999}\natexlab{}.
\newblock \showarticletitle{The cross-entropy method for combinatorial and
  continuous optimization}.
\newblock \bibinfo{journal}{\emph{Methodology and computing in applied
  probability}} \bibinfo{volume}{1}, \bibinfo{number}{2}
  (\bibinfo{year}{1999}), \bibinfo{pages}{127--190}.
\newblock


\bibitem[See et~al\mbox{.}(2017)]%
        {see2017get}
\bibfield{author}{\bibinfo{person}{Abigail See}, \bibinfo{person}{Peter~J Liu},
  {and} \bibinfo{person}{Christopher~D Manning}.}
  \bibinfo{year}{2017}\natexlab{}.
\newblock \showarticletitle{Get to the point: Summarization with
  pointer-generator networks}.
\newblock \bibinfo{journal}{\emph{arXiv preprint arXiv:1704.04368}}
  (\bibinfo{year}{2017}).
\newblock


\bibitem[Shariffdeen et~al\mbox{.}(2021)]%
        {shariffdeen2021concolic}
\bibfield{author}{\bibinfo{person}{Ridwan Shariffdeen}, \bibinfo{person}{Yannic
  Noller}, \bibinfo{person}{Lars Grunske}, {and} \bibinfo{person}{Abhik
  Roychoudhury}.} \bibinfo{year}{2021}\natexlab{}.
\newblock \showarticletitle{Concolic program repair}. In
  \bibinfo{booktitle}{\emph{Proceedings of the 42nd ACM SIGPLAN International
  Conference on Programming Language Design and Implementation}}.
  \bibinfo{pages}{390--405}.
\newblock


\bibitem[Shaw et~al\mbox{.}(2018)]%
        {shaw2018self}
\bibfield{author}{\bibinfo{person}{Peter Shaw}, \bibinfo{person}{Jakob
  Uszkoreit}, {and} \bibinfo{person}{Ashish Vaswani}.}
  \bibinfo{year}{2018}\natexlab{}.
\newblock \showarticletitle{Self-Attention with Relative Position
  Representations}. In \bibinfo{booktitle}{\emph{Proceedings of the 2018
  Conference of the North American Chapter of the Association for Computational
  Linguistics: Human Language Technologies, NAACL-HLT, New Orleans, Louisiana,
  USA, June 1-6, 2018, Volume 2 (Short Papers)}},
  \bibfield{editor}{\bibinfo{person}{Marilyn~A. Walker}, \bibinfo{person}{Heng
  Ji}, {and} \bibinfo{person}{Amanda Stent}} (Eds.).
  \bibinfo{publisher}{Association for Computational Linguistics},
  \bibinfo{pages}{464--468}.
\newblock
\urldef\tempurl%
\url{https://doi.org/10.18653/v1/n18-2074}
\showDOI{\tempurl}


\bibitem[Shi et~al\mbox{.}(2021)]%
        {shi2021cast}
\bibfield{author}{\bibinfo{person}{Ensheng Shi}, \bibinfo{person}{Yanlin Wang},
  \bibinfo{person}{Lun Du}, \bibinfo{person}{Hongyu Zhang},
  \bibinfo{person}{Shi Han}, \bibinfo{person}{Dongmei Zhang}, {and}
  \bibinfo{person}{Hongbin Sun}.} \bibinfo{year}{2021}\natexlab{}.
\newblock \showarticletitle{{CAST:} Enhancing Code Summarization with
  Hierarchical Splitting and Reconstruction of Abstract Syntax Trees}. In
  \bibinfo{booktitle}{\emph{Proceedings of the 2021 Conference on Empirical
  Methods in Natural Language Processing, {EMNLP} 2021, Virtual Event / Punta
  Cana, Dominican Republic, 7-11 November, 2021}},
  \bibfield{editor}{\bibinfo{person}{Marie{-}Francine Moens},
  \bibinfo{person}{Xuanjing Huang}, \bibinfo{person}{Lucia Specia}, {and}
  \bibinfo{person}{Scott~Wen{-}tau Yih}} (Eds.).
  \bibinfo{publisher}{Association for Computational Linguistics},
  \bibinfo{pages}{4053--4062}.
\newblock
\urldef\tempurl%
\url{https://doi.org/10.18653/v1/2021.emnlp-main.332}
\showDOI{\tempurl}


\bibitem[Svyatkovskiy et~al\mbox{.}(2019)]%
        {svyatkovskiy2019pythia}
\bibfield{author}{\bibinfo{person}{Alexey Svyatkovskiy}, \bibinfo{person}{Ying
  Zhao}, \bibinfo{person}{Shengyu Fu}, {and} \bibinfo{person}{Neel
  Sundaresan}.} \bibinfo{year}{2019}\natexlab{}.
\newblock \showarticletitle{Pythia: Ai-assisted code completion system}. In
  \bibinfo{booktitle}{\emph{Proceedings of the 25th ACM SIGKDD International
  Conference on Knowledge Discovery \& Data Mining}}.
  \bibinfo{pages}{2727--2735}.
\newblock


\bibitem[Tang et~al\mbox{.}(2021)]%
        {tang2021moto}
\bibfield{author}{\bibinfo{person}{Xunzhu Tang}, \bibinfo{person}{Rujie Zhu},
  \bibinfo{person}{Tiezhu Sun}, {and} \bibinfo{person}{Shi Wang}.}
  \bibinfo{year}{2021}\natexlab{}.
\newblock \showarticletitle{Moto: Enhancing Embedding with Multiple Joint
  Factors for Chinese Text Classification}. In \bibinfo{booktitle}{\emph{2020
  25th International Conference on Pattern Recognition (ICPR)}}. IEEE,
  \bibinfo{pages}{2882--2888}.
\newblock


\bibitem[Thunes(2013)]%
        {thunes2013javalang}
\bibfield{author}{\bibinfo{person}{Chris Thunes}.}
  \bibinfo{year}{2013}\natexlab{}.
\newblock \bibinfo{title}{javalang: pure Python Java parser and tools}.
\newblock
\newblock


\bibitem[Tian et~al\mbox{.}(2022a)]%
        {tian2022predicting}
\bibfield{author}{\bibinfo{person}{Haoye Tian}, \bibinfo{person}{Yinghua Li},
  \bibinfo{person}{Weiguo Pian}, \bibinfo{person}{Abdoul~Kader Kabore},
  \bibinfo{person}{Kui Liu}, \bibinfo{person}{Andrew Habib},
  \bibinfo{person}{Jacques Klein}, {and} \bibinfo{person}{Tegawend{\'e}~F
  Bissyand{\'e}}.} \bibinfo{year}{2022}\natexlab{a}.
\newblock \showarticletitle{Predicting Patch Correctness Based on the
  Similarity of Failing Test Cases}.
\newblock \bibinfo{journal}{\emph{ACM Transactions on Software Engineering and
  Methodology}} (\bibinfo{year}{2022}).
\newblock


\bibitem[Tian et~al\mbox{.}(2020)]%
        {tian2020evaluating}
\bibfield{author}{\bibinfo{person}{Haoye Tian}, \bibinfo{person}{Kui Liu},
  \bibinfo{person}{Abdoul~Kader Kabor{\'e}}, \bibinfo{person}{Anil Koyuncu},
  \bibinfo{person}{Li Li}, \bibinfo{person}{Jacques Klein}, {and}
  \bibinfo{person}{Tegawend{\'e}~F Bissyand{\'e}}.}
  \bibinfo{year}{2020}\natexlab{}.
\newblock \showarticletitle{Evaluating representation learning of code changes
  for predicting patch correctness in program repair}. In
  \bibinfo{booktitle}{\emph{2020 35th IEEE/ACM International Conference on
  Automated Software Engineering (ASE)}}. IEEE, \bibinfo{pages}{981--992}.
\newblock


\bibitem[Tian et~al\mbox{.}(2022b)]%
        {tian2022best}
\bibfield{author}{\bibinfo{person}{Haoye Tian}, \bibinfo{person}{Kui Liu},
  \bibinfo{person}{Yinghua Li}, \bibinfo{person}{Abdoul~Kader Kabor{\'e}},
  \bibinfo{person}{Anil Koyuncu}, \bibinfo{person}{Andrew Habib},
  \bibinfo{person}{Li Li}, \bibinfo{person}{Junhao Wen},
  \bibinfo{person}{Jacques Klein}, {and} \bibinfo{person}{Tegawend{\'e}~F
  Bissyand{\'e}}.} \bibinfo{year}{2022}\natexlab{b}.
\newblock \showarticletitle{The Best of Both Worlds: Combining Learned
  Embeddings with Engineered Features for Accurate Prediction of Correct
  Patches}.
\newblock \bibinfo{journal}{\emph{arXiv preprint arXiv:2203.08912}}
  (\bibinfo{year}{2022}).
\newblock


\bibitem[Tian et~al\mbox{.}(2022c)]%
        {tian2022change}
\bibfield{author}{\bibinfo{person}{Haoye Tian}, \bibinfo{person}{Xunzhu Tang},
  \bibinfo{person}{Andrew Habib}, \bibinfo{person}{Shangwen Wang},
  \bibinfo{person}{Kui Liu}, \bibinfo{person}{Xin Xia},
  \bibinfo{person}{Jacques Klein}, {and} \bibinfo{person}{Tegawend{\'e}~F
  Bissyand{\'e}}.} \bibinfo{year}{2022}\natexlab{c}.
\newblock \showarticletitle{Is this Change the Answer to that Problem?
  Correlating Descriptions of Bug and Code Changes for Evaluating Patch
  Correctness}.
\newblock \bibinfo{journal}{\emph{arXiv preprint arXiv:2208.04125}}
  (\bibinfo{year}{2022}).
\newblock


\bibitem[Van~der Maaten and Hinton(2008)]%
        {van2008visualizing}
\bibfield{author}{\bibinfo{person}{Laurens Van~der Maaten} {and}
  \bibinfo{person}{Geoffrey Hinton}.} \bibinfo{year}{2008}\natexlab{}.
\newblock \showarticletitle{Visualizing data using t-SNE.}
\newblock \bibinfo{journal}{\emph{Journal of machine learning research}}
  \bibinfo{volume}{9}, \bibinfo{number}{11} (\bibinfo{year}{2008}).
\newblock


\bibitem[Vijayakumar et~al\mbox{.}(2016)]%
        {vijayakumar2016diverse}
\bibfield{author}{\bibinfo{person}{Ashwin~K Vijayakumar},
  \bibinfo{person}{Michael Cogswell}, \bibinfo{person}{Ramprasath~R Selvaraju},
  \bibinfo{person}{Qing Sun}, \bibinfo{person}{Stefan Lee},
  \bibinfo{person}{David Crandall}, {and} \bibinfo{person}{Dhruv Batra}.}
  \bibinfo{year}{2016}\natexlab{}.
\newblock \showarticletitle{Diverse beam search: Decoding diverse solutions
  from neural sequence models}.
\newblock \bibinfo{journal}{\emph{arXiv preprint arXiv:1610.02424}}
  (\bibinfo{year}{2016}).
\newblock


\bibitem[Wang et~al\mbox{.}(2021b)]%
        {wang2021context}
\bibfield{author}{\bibinfo{person}{Haoye Wang}, \bibinfo{person}{Xin Xia},
  \bibinfo{person}{David Lo}, \bibinfo{person}{Qiang He},
  \bibinfo{person}{Xinyu Wang}, {and} \bibinfo{person}{John Grundy}.}
  \bibinfo{year}{2021}\natexlab{b}.
\newblock \showarticletitle{Context-aware retrieval-based deep commit message
  Generation}.
\newblock \bibinfo{journal}{\emph{ACM Transactions on Software Engineering and
  Methodology (TOSEM)}} \bibinfo{volume}{30}, \bibinfo{number}{4}
  (\bibinfo{year}{2021}), \bibinfo{pages}{1--30}.
\newblock


\bibitem[Wang et~al\mbox{.}(2017)]%
        {wang2017dynamic}
\bibfield{author}{\bibinfo{person}{Ke Wang}, \bibinfo{person}{Rishabh Singh},
  {and} \bibinfo{person}{Zhendong Su}.} \bibinfo{year}{2017}\natexlab{}.
\newblock \showarticletitle{Dynamic neural program embedding for program
  repair}.
\newblock \bibinfo{journal}{\emph{arXiv preprint arXiv:1711.07163}}
  (\bibinfo{year}{2017}).
\newblock


\bibitem[Wang et~al\mbox{.}(2019)]%
        {dgl}
\bibfield{author}{\bibinfo{person}{Minjie Wang}, \bibinfo{person}{Da Zheng},
  \bibinfo{person}{Zihao Ye}, \bibinfo{person}{Quan Gan},
  \bibinfo{person}{Mufei Li}, \bibinfo{person}{Xiang Song},
  \bibinfo{person}{Jinjing Zhou}, \bibinfo{person}{Chao Ma},
  \bibinfo{person}{Lingfan Yu}, \bibinfo{person}{Yu Gai}, {et~al\mbox{.}}}
  \bibinfo{year}{2019}\natexlab{}.
\newblock \showarticletitle{Deep graph library: A graph-centric,
  highly-performant package for graph neural networks}.
\newblock \bibinfo{journal}{\emph{arXiv preprint arXiv:1909.01315}}
  (\bibinfo{year}{2019}).
\newblock


\bibitem[Wang et~al\mbox{.}(2022a)]%
        {wang2022hienet}
\bibfield{author}{\bibinfo{person}{Shi Wang}, \bibinfo{person}{Daniel Tang},
  \bibinfo{person}{Luchen Zhang}, \bibinfo{person}{Huilin Li}, {and}
  \bibinfo{person}{Ding Han}.} \bibinfo{year}{2022}\natexlab{a}.
\newblock \showarticletitle{Hienet: Bidirectional hierarchy framework for
  automated icd coding}. In \bibinfo{booktitle}{\emph{International Conference
  on Database Systems for Advanced Applications}}. Springer,
  \bibinfo{pages}{523--539}.
\newblock


\bibitem[Wang et~al\mbox{.}(2022b)]%
        {DBLP:conf/dasfaa/WangTZLH22}
\bibfield{author}{\bibinfo{person}{Shi Wang}, \bibinfo{person}{Daniel Tang},
  \bibinfo{person}{Luchen Zhang}, \bibinfo{person}{Huilin Li}, {and}
  \bibinfo{person}{Ding Han}.} \bibinfo{year}{2022}\natexlab{b}.
\newblock \showarticletitle{HieNet: Bidirectional Hierarchy Framework for
  Automated {ICD} Coding}. In \bibinfo{booktitle}{\emph{Database Systems for
  Advanced Applications - 27th International Conference, {DASFAA} 2022, Virtual
  Event, April 11-14, 2022, Proceedings, Part {II}}}
  \emph{(\bibinfo{series}{Lecture Notes in Computer Science},
  Vol.~\bibinfo{volume}{13246})}, \bibfield{editor}{\bibinfo{person}{Arnab
  Bhattacharya}, \bibinfo{person}{Janice Lee}, \bibinfo{person}{Mong Li},
  \bibinfo{person}{Divyakant Agrawal}, \bibinfo{person}{P.~Krishna Reddy},
  \bibinfo{person}{Mukesh~K. Mohania}, \bibinfo{person}{Anirban Mondal},
  \bibinfo{person}{Vikram Goyal}, {and} \bibinfo{person}{Rage~Uday Kiran}}
  (Eds.). \bibinfo{publisher}{Springer}, \bibinfo{pages}{523--539}.
\newblock
\urldef\tempurl%
\url{https://doi.org/10.1007/978-3-031-00126-0\_38}
\showDOI{\tempurl}


\bibitem[Wang et~al\mbox{.}(2021a)]%
        {DBLP:conf/emnlp/0034WJH21}
\bibfield{author}{\bibinfo{person}{Yue Wang}, \bibinfo{person}{Weishi Wang},
  \bibinfo{person}{Shafiq~R. Joty}, {and} \bibinfo{person}{Steven C.~H. Hoi}.}
  \bibinfo{year}{2021}\natexlab{a}.
\newblock \showarticletitle{CodeT5: Identifier-aware Unified Pre-trained
  Encoder-Decoder Models for Code Understanding and Generation}. In
  \bibinfo{booktitle}{\emph{Proceedings of the 2021 Conference on Empirical
  Methods in Natural Language Processing, {EMNLP} 2021, Virtual Event / Punta
  Cana, Dominican Republic, 7-11 November, 2021}},
  \bibfield{editor}{\bibinfo{person}{Marie{-}Francine Moens},
  \bibinfo{person}{Xuanjing Huang}, \bibinfo{person}{Lucia Specia}, {and}
  \bibinfo{person}{Scott~Wen{-}tau Yih}} (Eds.).
  \bibinfo{publisher}{Association for Computational Linguistics},
  \bibinfo{pages}{8696--8708}.
\newblock
\urldef\tempurl%
\url{https://doi.org/10.18653/v1/2021.emnlp-main.685}
\showDOI{\tempurl}


\bibitem[Xu et~al\mbox{.}(2019)]%
        {xu2019commit}
\bibfield{author}{\bibinfo{person}{Shengbin Xu}, \bibinfo{person}{Yuan Yao},
  \bibinfo{person}{Feng Xu}, \bibinfo{person}{Tianxiao Gu},
  \bibinfo{person}{Hanghang Tong}, {and} \bibinfo{person}{Jian Lu}.}
  \bibinfo{year}{2019}\natexlab{}.
\newblock \showarticletitle{Commit message generation for source code changes}.
  In \bibinfo{booktitle}{\emph{IJCAI}}.
\newblock


\bibitem[Yin et~al\mbox{.}(2019)]%
        {yin2018learning}
\bibfield{author}{\bibinfo{person}{Pengcheng Yin}, \bibinfo{person}{Graham
  Neubig}, \bibinfo{person}{Miltiadis Allamanis}, \bibinfo{person}{Marc
  Brockschmidt}, {and} \bibinfo{person}{Alexander~L. Gaunt}.}
  \bibinfo{year}{2019}\natexlab{}.
\newblock \showarticletitle{Learning to Represent Edits}. In
  \bibinfo{booktitle}{\emph{International Conference on Learning
  Representations}}.
\newblock
\urldef\tempurl%
\url{https://openreview.net/forum?id=BJl6AjC5F7}
\showURL{%
\tempurl}


\bibitem[Zhang et~al\mbox{.}(2019d)]%
        {zhang2019novel}
\bibfield{author}{\bibinfo{person}{Jian Zhang}, \bibinfo{person}{Xu Wang},
  \bibinfo{person}{Hongyu Zhang}, \bibinfo{person}{Hailong Sun},
  \bibinfo{person}{Kaixuan Wang}, {and} \bibinfo{person}{Xudong Liu}.}
  \bibinfo{year}{2019}\natexlab{d}.
\newblock \showarticletitle{A novel neural source code representation based on
  abstract syntax tree}. In \bibinfo{booktitle}{\emph{2019 IEEE/ACM 41st
  International Conference on Software Engineering (ICSE)}}. IEEE,
  \bibinfo{pages}{783--794}.
\newblock


\bibitem[Zhang et~al\mbox{.}(2019c)]%
        {zhang2019graph}
\bibfield{author}{\bibinfo{person}{Si Zhang}, \bibinfo{person}{Hanghang Tong},
  \bibinfo{person}{Jiejun Xu}, {and} \bibinfo{person}{Ross Maciejewski}.}
  \bibinfo{year}{2019}\natexlab{c}.
\newblock \showarticletitle{Graph convolutional networks: a comprehensive
  review}.
\newblock \bibinfo{journal}{\emph{Computational Social Networks}}
  \bibinfo{volume}{6}, \bibinfo{number}{1} (\bibinfo{year}{2019}),
  \bibinfo{pages}{1--23}.
\newblock


\bibitem[Zhang et~al\mbox{.}(2019a)]%
        {zhang2019ernie}
\bibfield{author}{\bibinfo{person}{Zhengyan Zhang}, \bibinfo{person}{Xu Han},
  \bibinfo{person}{Zhiyuan Liu}, \bibinfo{person}{Xin Jiang},
  \bibinfo{person}{Maosong Sun}, {and} \bibinfo{person}{Qun Liu}.}
  \bibinfo{year}{2019}\natexlab{a}.
\newblock \showarticletitle{ERNIE: Enhanced language representation with
  informative entities}.
\newblock \bibinfo{journal}{\emph{arXiv preprint arXiv:1905.07129}}
  (\bibinfo{year}{2019}).
\newblock


\bibitem[Zhang et~al\mbox{.}(2019b)]%
        {zhang-etal-2019-ernie}
\bibfield{author}{\bibinfo{person}{Zhengyan Zhang}, \bibinfo{person}{Xu Han},
  \bibinfo{person}{Zhiyuan Liu}, \bibinfo{person}{Xin Jiang},
  \bibinfo{person}{Maosong Sun}, {and} \bibinfo{person}{Qun Liu}.}
  \bibinfo{year}{2019}\natexlab{b}.
\newblock \showarticletitle{{ERNIE}: Enhanced Language Representation with
  Informative Entities}. In \bibinfo{booktitle}{\emph{Proceedings of the 57th
  Annual Meeting of the Association for Computational Linguistics}}.
  \bibinfo{publisher}{Association for Computational Linguistics},
  \bibinfo{address}{Florence, Italy}, \bibinfo{pages}{1441--1451}.
\newblock
\urldef\tempurl%
\url{https://doi.org/10.18653/v1/P19-1139}
\showDOI{\tempurl}


\end{thebibliography}

\end{document}